\documentclass{aa}

\usepackage[utf8]{inputenc}
\usepackage[T1]{fontenc}
\usepackage{graphicx}
\usepackage{xcolor}
\usepackage{amsmath}
\usepackage{txfonts}
\usepackage{booktabs}
\usepackage{array}
\usepackage{longtable}
\usepackage{pdflscape}
\usepackage{float}
\usepackage{natbib}
\usepackage{url}

\newcommand{\code}[1]{\texttt{\detokenize{#1}}}

\newcommand{\dscb}{\ensuremath{\Delta S_{\rm CB}}}

\providecommand{\apj}{ApJ}
\providecommand{\apjl}{ApJL}
\providecommand{\apjs}{ApJS}
\providecommand{\aap}{A\&A}
\providecommand{\mnras}{MNRAS}
\providecommand{\nat}{Nature}
\providecommand{\apss}{Ap\&SS}

\begin{document}

\title{Reliability of Fermi/GBM spectral-catalogue parameters for gamma-ray burst population studies}
\titlerunning{Reliability of Fermi/GBM spectral parameters}

\author{I. Horvath\corrauth{horvath.istvan@uni-nke.hu}}
\authorrunning{I. Horvath}
\institute{National University of Public Service, Budapest, Hungary}
\date{} 

\abstract
{The Fermi Gamma-ray Burst Monitor (GBM) spectral catalogue contains prompt-emission measurements whose interpretation is complicated by model-dependent uncertainties, distinct peak-flux (PFLX) and fluence (FLNC) measurements, different PFLX integration times, and changes in catalogue processing.}
{We determine which quantities represent spectral shape and which trace brightness, parameter precision, or measurement conditions.}
{We analysed 3337 of 3378 GRBs containing the required spectral fields. We identified a transition at \code{GRB180712358}, defined two periods, and treated the 64 and nominal 1024 ms PFLX integrations as distinct measurements. We used rank-based comparisons and regressions, catalogue model-selection rules, and repeated cross-validated duration classification. Short--long comparisons used FLNC quantities only.}
{The transition coincides with changes in catalogue version, statistic type, and fit-field behaviour. In the nominal 1024 ms sample, all ten COMP A--B effect estimates satisfy $|r_{\rm rb}|<0.10$. In the 64 ms sample, COMP shape parameters remain similar, but amplitudes, fluxes, and fluences are higher in the later period; the origin of this difference is unknown. Within the earlier period, brighter bursts have smaller parameter errors but larger reported fit statistics and stronger evidence for curvature. The FLNC PLAW index is an effective broad-band hardness measure. BAND $\alpha$ and $E_{\rm peak}$ are usable when their own errors are controlled, whereas $\beta$ is often poorly constrained. Adding PLAW or BAND to COMP provides little duration-classification information in matched samples.}
{The A--B division is best interpreted as a catalogue-processing transition rather than population evolution. PFLX measurements at different integration times should not be pooled directly, whereas FLNC provides a common basis for short--long comparisons. Low-energy indices and peak energies with controlled uncertainties are suitable spectral-shape variables. Fit statistics, parameter errors, catalogue metadata, and model labels should be treated as measurement indicators; correlated model parameters should not be treated as independent.}


\maketitle
\nolinenumbers

\section{Introduction}\label{sec:intro}
Gamma-ray bursts (GRBs) are commonly divided into short and long events by the observed duration $T_{90}$, with the traditional boundary near 2 s \citep{maz81,nor84,kou93}. This division is useful but is not a complete physical classification. Duration depends on detector band, sensitivity, background treatment, trigger logic, and prompt-emission morphology \citep{kos96,hak00,band03,qin13}. Some physically unusual events cross the conventional duration boundary, and recent duration-free or progenitor-based analyses also show that duration alone does not define a unique physical class \citep{hth18,Zhu2024,Chen2025DeepLearning,Wei2026AnomalousGRB,Zhang2026SpectralHardness}. Hence, spectral measurements provide an important complementary description of the prompt emission.

The number of observational GRB groups has remained method dependent. Evidence for an intermediate or third component has been reported from univariate duration fits, duration--hardness analyses, and multivariate studies of samples from the Burst and Transient Source Experiment (BATSE), Swift, BeppoSAX, and Fermi \citep{hor98,muk98,hor02,bala01,hak03,hor06,chat07,hor08,hor09,hor10,veres10,kb12,zito15,yzj16,hth18,hoi19ApS,2022GalaxSalmon,Bhave2022TwoDimensional}. Other studies favour two dominant components, or show that the preference between two and three components changes with distributional assumptions, sample size, and the representation of the prompt emission \citep{3gbm16,tarno15AA,tarno16MNRAS,kulk17,Jespersen2020,Salmon2022FirstSecond,2022MNRAS517.5770Z,Steinhardt2023,Chen2024FermiSpectralML,Zhu2024,Chen2025DeepLearning,Modak2025BinaryClustering,Tsvetkova2025GRBTaxonomy}. Some studies have also obtained four- and five-component solutions using Gaussian, nested-Gaussian, multivariate-$t$, and nonlinear machine-learning representations \citep{chat17,chat18,acuneryde18,Mehta2024GRBDiversity,Dimple2024FermiGBMML}. Reanalyses have shown, however, that some additional mixture components can be brightness subdivisions of broader duration--hardness classes rather than distinct source populations \citep{Toth2019BATSEGMM,hoi19ApS}. These results show that intrinsic spectral information must be separated from brightness, parameter errors, catalogue processing, and differences in measurement quality before the multivariate groups are interpreted.

The Fermi Gamma-ray Burst Monitor (GBM) provides a large set of prompt-emission measurements. Its instrument calibration, burst catalogues, and spectral catalogues give durations, fluences, peak fluxes, empirical spectral-model parameters, uncertainties, model labels, and fit-statistic fields \citep{2009ApJ...702..791M,fermi_gbm1,fermi_gbm2,fermi4,fermi2,gold12,gru14,3gbm16,vonKienlin2020GBM4,Poolakkil2021GBM10yr,yu_evolution}. The BATSE 5B spectral catalogue is the complete-mission predecessor based on the same empirical model family and provides a reference for comparison with the GBM results \citep{Goldstein2013BATSE5B}. Many of the catalogue variables are strongly correlated. Brightness is measured in several bands and time scales, and the same spectrum is described by empirical models with different numbers of parameters and different parameter errors. Earlier principal component analysis (PCA) and multivariate studies found strong correlations and overlap in the information carried by catalogue variables \citep{bag98,racz_an_fermi_2018,hoi19ApS,Chen2024FermiSpectralML}. In model-based clustering, irrelevant or duplicated variables can change the classification; removing variables that contain information on clustering can also change the groups found \citep{RafteryDean2006VariableSelection,chat18}.

The catalogue quantities do not all have the same measurement definition. Fluence-spectrum (FLNC) fits integrate over much of the burst, whereas peak-flux-spectrum (PFLX) fits describe a short interval around the brightest phase. The standard PFLX interval is nominal 64 ms for short bursts and nominal 1024 ms for long bursts \citep{gold12,gru14,Poolakkil2021GBM10yr}. Time-resolved and FLNC--PFLX comparisons show that the brightest phase is often spectrally harder than the time-integrated emission \citep{Kumar8_norris,2001ApJ...548..770B,Kumar11_shortgrbs,yu_evolution,2018Rac}.

The catalogue uses four empirical spectral models: a power law (PLAW), a cutoff or Comptonized form (COMP), the BAND function, and a smoothly broken power law (SBPL) \citep{band,comptonized1,smothly,Kumar16,2006ApJS..166..298K,gold12,gru14,Poolakkil2021GBM10yr}. For a faint spectrum, a simple model can be preferred because additional curvature is not constrained, whereas a bright spectrum can reject the same function even when its fitted parameters have small errors. Even when it is not measured reliably, a parameter may be listed in the catalogue, and $E_{\rm peak}$ can also be shifted by the adopted spectral form or by analysis choices not included in the quoted counting-statistical uncertainty \citep{Collazzi2011EpeakErrors,2016ApJ...821...12P,Li2023SpectralEnergy}. These points are especially important for the BAND high-energy index and for SBPL parameters in faint spectra or spectra with low signal-to-noise ratio \citep{2018Rac,hoi19ApS,Poolakkil2021GBM10yr}.

We examine the catalogue structure and the reported fit-statistic behaviour of the spectral models. We locate the change in catalogue processing and test the stability of the central spectral parameters on its two sides. The 64 and 1024 ms PFLX measurements are treated separately when studying the relation of brightness to parameter precision and the reported COMP statistic within each fixed measurement group. We also identify the COMP, PLAW, and BAND quantities that can be used as spectral-shape parameters in population analysis. Population clustering and principal-component population analysis are not performed in this paper.

This is the first of two papers. Here we establish the measurement and variable-selection rules. In the second paper, we will apply these rules to compare the sets of variables defined here and to investigate possible subclasses in the Fermi/GBM population.

Section~\ref{sec:variables} describes the catalogue measurements and spectral-model blocks. Section~\ref{sec:data} defines the 3337-GRB analysis sample and the initial model comparisons. Section~\ref{sec:groups} identifies the catalogue transition and the valid PFLX comparison groups. Section~\ref{sec:compelemzes} compares the COMP physical parameters on the two sides of the catalogue transition, while Sect.~\ref{sec:compfitquality} examines how brightness, parameter errors, and the reported COMP statistic are related within each group. Sections~\ref{sec:plaw} and \ref{sec:band} establish the roles and limitations of PLAW and BAND, followed by the discussion and conclusions.

\section{Fermi/GBM spectral catalogue and analysis variables}\label{sec:variables}
On 2026 April 7, the Fermi GBM catalogue contained 4248 GRBs. The most recent event was GRB260407383; the last burst with spectral-fit fields available in the analysed catalogue data was GRB241030242, the 3378th event in time order. Forty-one records lacked at least one spectral field required by the present analyses; the complete spectral-parameter sample therefore contains 3337 GRBs.

The catalogue contains both event-level and model-dependent measurements. Event-level quantities include $T_{50}$, $T_{90}$, fluences, and peak fluxes in several energy bands and time intervals. The catalogue fits spectral quantities separately to FLNC and PFLX data selections \citep{gold12,gru14,Poolakkil2021GBM10yr}. FLNC usually contains more photons and averages over spectral evolution; PFLX samples the brightest phase and is often harder, as found in time-resolved and catalogue-level comparisons \citep{Kumar8_norris,Kumar11_shortgrbs,yu_evolution,2018Rac,racz_an_fermi_2018}. The two blocks are physically related, but they should not be treated as the same measurement.

PLAW contains an amplitude and one photon index. COMP adds a cutoff and a peak-energy scale. BAND contains a low-energy index $\alpha$, a high-energy index $\beta$, and a characteristic energy usually represented by $E_{\rm peak}$. SBPL uses two power-law indices and a smooth break \citep{band,comptonized1,smothly}. The models have different numbers of parameters, and their parameters are not measured with the same accuracy. The official spectral catalogues therefore combine fit-statistic improvement with parameter-quality criteria instead of selecting the numerically smallest statistic alone \citep{gold12,gru14,Poolakkil2021GBM10yr}; a lower statistic for a more complex model does not by itself imply a physically useful solution. Likelihood-ratio comparisons also require special care when a null value lies on the boundary of the parameter space \citep{Protassov2002LRT}.

Following the treatment used in earlier multivariate analyses of Fermi GBM data, variables spanning large dynamical ranges are expressed on a base-10 logarithmic scale \citep{hoi19ApS}; spectral indices remain on their original scale. Related flux and fluence measurements are used as brightness variables and not as independent spectral-shape variables, consistent with earlier PCA and multivariate GRB studies \citep{bag98,2018Rac,hoi19ApS} and with general principles for selecting variables \citep{RafteryDean2006VariableSelection}.

Fit statistics, parameter uncertainties, catalogue version and statistic type, catalogue period, and PFLX integration duration are used to examine measurement, fit-statistic, and processing effects. They are not treated as intrinsic GRB properties.

\section{Sample construction and initial spectral-model comparisons}\label{sec:data}
\subsection{Construction of the analysis sample}\label{subsec:sampleconstruction}
We use the complete 3337-GRB spectral-parameter sample whenever the required model fields are present. Model-specific quality conditions define smaller subsets in the individual analyses, but no single global quality cut is imposed across the COMP, PLAW, and BAND chapters because these models contain different parameters that are not measured with the same accuracy. We apply only one row selection to the full catalogue: 41 of the 3378 spectral-fit records are excluded because at least one required spectral field is absent, leaving 3337 GRBs.

We also checked the brightness of short GRBs using the catalogue data. We selected bursts with $T_{90}<2$ s within the period of this study. We used the event-level 64 ms peak flux (\code{FLUX\_64}) and checked whether a 64 ms COMP PFLX fit with positive amplitude was available. We applied the same conditions in A and B.

\subsection{Raw model-rank comparisons}\label{subsec:modelchecks}
The catalogue has two different fields. The \code{redchisq} field gives the reduced chi-squared value at the fitted parameters. The \code{redfitstat} field gives the reduced value of the statistic used for the fit. The \code{statistic} field identifies this statistic \citep{gold12}.\footnote{\url{https://heasarc.gsfc.nasa.gov/W3Browse/fermi/fermigbrst.html}} We retain this distinction throughout. Our \code{reducechisq} notation refers to the reported \code{redchisq} field and does not imply a common statistical interpretation across catalogue periods. In particular, the equality of \code{redchisq} and \code{redfitstat} in A, but generally not in B, requires caution when comparing their numerical values (Sect.~\ref{app:compfitfield}). A smaller reported value alone does not establish better parameter constraints.

For the four raw ranking results summarized in Appendix~\ref{app:rankchecks}, we use all 3378 records with spectral-fit products; all subsequent analyses use the 3337 records containing every field required by the corresponding analysis. Tables~\ref{tab:flnc_legroszab_chi2model_3378} and \ref{tab:pflx_legroszab_chi2model_3378} give the model with the largest reported reduced-statistic value on each side. We find that PLAW is most often the poorest numerical fit, particularly for FLNC, consistent with the established ordering from a simple power law to more flexible curved models \citep{2018Rac,racz_an_fermi_2018}. This descriptive ranking is not equivalent to the catalogue GOOD/BEST decision, which also depends on parameter quality and the defined fit-statistic improvements between nested model levels \citep{gold12,gru14,Poolakkil2021GBM10yr}. Section~\ref{sec:plaw} returns to this distinction directly.

Table~\ref{tab:biggest_chi2_flnc_pflx_3378} shows the joint distribution of the worst-ranked FLNC and PFLX models.

After removing PLAW from the raw-rank comparison, Table~\ref{tab:noplaw-3x3} compares the curved model with the smallest reported reduced-statistic value on the FLNC and PFLX sides.

For each GRB, COMP, BAND, and SBPL were also ranked separately on the FLNC and PFLX sides. Appendix~\ref{app:rankchecks} gives the full distribution of the differences between these ranks for the 3337-GRB sample as a supporting catalogue comparison. It is not used to select individual GRBs in the present analysis.

\section{Catalogue transition and comparison groups}\label{sec:groups}

We used the complete 3337-GRB spectral-parameter sample to study the catalogue transition. The comparison groups are defined by two catalogue features: a change with time in the reported fit-statistic fields and the duration of the peak-flux spectral integration interval. The sample is divided into four standard groups and one small residual group, and all 3337 GRBs are retained. These groups are used only for comparison and are not interpreted as possible astrophysical populations.

\subsection{Identification of the catalogue transition}\label{subsec:catalogtransition}

We find a clear transition in the catalogue metadata at \code{GRB180712358}, on 2018 July 12. We checked the eight available model- and spectrum-specific statistic fields: PFLX and FLNC for PLAW, COMP, BAND, and SBPL. Before the boundary, all eight statistic fields are uniformly recorded as \code{Castor C-STAT}. Beginning with \code{GRB180712358}, the catalogue changes to \code{PSTAT}: all non-missing COMP, BAND, and SBPL statistic fields are PSTAT, while only five GRBs retain Castor C-STAT in one or both PLAW statistic fields. Hence, the statistic labels themselves identify the catalogue transition without the need for a separately estimated change point.

\begin{figure}[htbp]
\centering
\includegraphics[width=0.95\columnwidth]{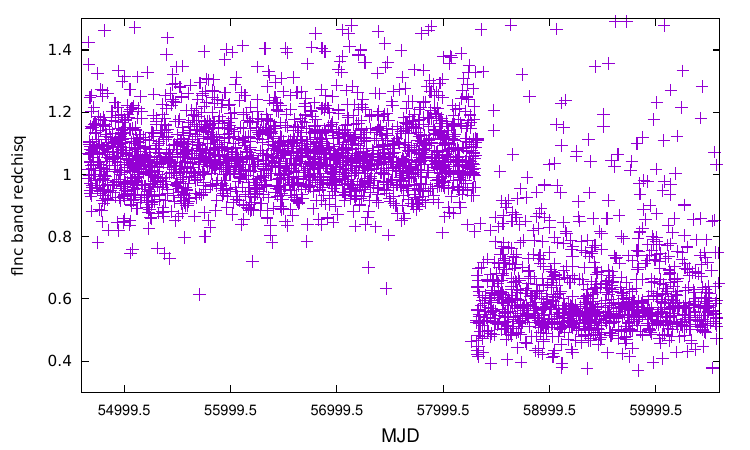}
\caption{Time-ordered reported FLNC reduced-statistic values for the BAND fits in the 3337-GRB sample. A clear change occurs at \code{GRB180712358} (MJD 58311.358), the boundary between the pre-transition B and post-transition A catalogue periods.}
\label{fig:bandflncbreak}
\end{figure}

The spectral-catalogue version changes at the same boundary. In the 3337-GRB analysis sample, the 2244 earlier rows have \code{scatalog}=3 or 4, whereas all 1093 rows beginning with \code{GRB180712358}, at Modified Julian Date (MJD) 58311.358, have \code{scatalog}=5. The reported BAND FLNC reduced-statistic values show the same transition, as illustrated in Fig.~\ref{fig:bandflncbreak}. We refer to the earlier interval as the pre-transition B period and the later interval as the post-transition A period; these are catalogue-processing periods rather than astrophysical GRB populations.

The Castor modification of the Cash statistic is the Poisson deviance, $C=2\sum_i[\mu_i-n_i+n_i\ln(n_i/\mu_i)]$, where $n_i$ and $\mu_i$ are the observed and expected source-plus-background counts in channel $i$; the logarithmic term is zero for $n_i=0$ \citep{1979ApJ...228..939C,Kaastra2017Cstat}. For a background treated as known, the XSPEC definition of PSTAT has the same form.\footnote{\url{https://heasarc.gsfc.nasa.gov/docs/software/xspec/manual/node393.html}} Thus, different names do not necessarily mean different distributions when the data and fitting methods are the same. However, the XSPEC definition does not tell us exactly how each GBM catalogue value was calculated or stored.

Even when the Poisson model is correct, the mean and variance of this statistic depend on the expected counts in each channel. Dividing by the degrees of freedom does not always give a mean of one \citep{Kaastra2017Cstat}. Different samples can therefore have different statistic distributions without a change of definition. Background treatment, channel selection, and numerical fitting procedures can introduce further differences. A positive constant normalization factor rescales the statistic and its differences between models, but leaves the exact best-fitting parameters unchanged and cannot by itself explain changes in fitted fluxes or spectral parameters.

We cannot conclude from the A--B statistic difference that the physical fit quality changed. We also do not use a common factor of two. The change in statistic names, catalogue version, and fit fields all occur at the same boundary. We use absolute statistic limits only in B. The A--B differences help us check whether the catalogue is uniform.

\subsection{Peak-flux spectral time scale and comparison criteria}\label{subsec:pflxscale}

We calculated the peak-flux spectral integration time from
\begin{equation}
\Delta t_{\rm pflx}=t_{\rm stop,pflx}-t_{\rm start,pflx},
\end{equation}
using \code{pflxspectrumstart} and \code{pflxspectrumstop}. The nominal PFLX integration times are 64 and 1024 ms. The intervals calculated from the catalogue are 0.064 s in both short groups and 1.024 s in longA. In longB, most are also 1.024 s (1800 of 1848), but the full range is 1.020--1.040 s. The former is normally used for bursts with $T_{90}<2$ s, whereas the latter is normally used for $T_{90}\geq2$ s, following the construction of the GBM spectral catalogues \citep{gold12,gru14,Poolakkil2021GBM10yr}. In A this rule is exact in the present sample: all 103 events fitted on 64 ms have $T_{90}<2$ s, and all 990 events fitted on 1024 ms have $T_{90}\geq2$ s. In B, 359 of the 375 events fitted on 64 ms have $T_{90}<2$ s, but 16 have $T_{90}\geq2$ s. Thus, the labels \code{short} and \code{long} refer primarily to the actual PFLX integration scale rather than to an independently imposed duration cut.

The nominal PFLX integration times differ by a factor of 16; hence, the two measurements cannot be treated as if they were obtained on the same time scale. We apply the following rules:
\begin{enumerate}
\item shortA is compared with shortB for PFLX quantities because both use the nominal 64 ms integration time;
\item longA is compared with longB for PFLX quantities because both use nominal 1024 ms;
\item short and long PFLX quantities are neither pooled nor entered into a common physical short--long test, and are not interpreted as measurements on one scale;
\item short--long and four-group physical comparisons are based on FLNC quantities only;
\end{enumerate}
Displaying separately fitted coefficients from the short and long PFLX analyses in a common table does not change this rule; the original PFLX observations and their likelihoods remain separate.

\subsection{The residual \code{unic} group}\label{subsec:unic}

\begin{figure}[htbp]
\centering
\includegraphics[width=0.90\columnwidth]{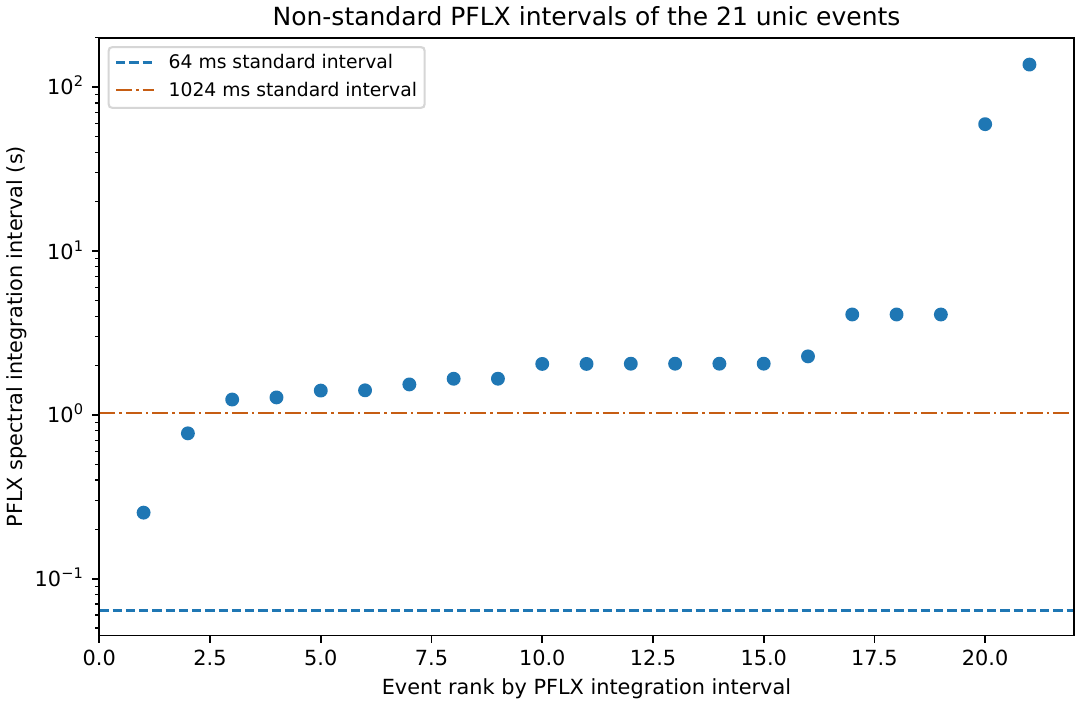}
\caption{PFLX spectral integration intervals of the 21 \code{unic} events. The dashed reference levels mark the standard 64 and 1024 ms intervals.}
\label{fig:unicintervals}
\end{figure}

The four standard groups contain 3316 GRBs. The other 21 B events form the \code{unic} group. Although all have $T_{90}>2$ s, their PFLX spectral intervals range from 0.253 to 137.220 s and do not correspond to either standard PFLX scale. They are therefore kept for completeness but excluded from all PFLX group comparisons. An FLNC-only sensitivity check in which the 21 events were temporarily added to longB changed the five central longA--longB FLNC rank effects by at most 0.0038. Omitting these events has a very small influence on the FLNC conclusions, whereas assigning them to either standard PFLX group would not be justified. Figure~\ref{fig:unicintervals} shows their non-standard PFLX intervals, and Table~\ref{tab:fivegroups} summarizes the resulting five catalogue groups.

\begin{table*}[htbp]
\centering
\scriptsize
\caption{Partition of the 3337-GRB spectral-parameter sample. The A period starts with \code{GRB180712358}; this boundary row belongs to A. The \code{short}/\code{long} labels are based on the actual PFLX spectral interval.}
\label{tab:fivegroups}
\begin{tabular}{lccrrrrl}
\toprule
Group & Period & $\Delta t_{\rm pflx}$ (s) & $N$ & $T_{90}<2$ s & $T_{90}\geq2$ s & Fraction of 3337 & GRB-name range \\
\midrule
\code{shortA} & A & 0.064 & 103 & 103 & 0 & 3.09\% & GRB180715755--GRB241016680 \\
\code{shortB} & B & 0.064 & 375 & 359 & 16 & 11.24\% & GRB080723913--GRB180703949 \\
\code{longA}  & A & 1.024 & 990 & 0 & 990 & 29.67\% & GRB180712358--GRB241030242 \\
\code{longB}  & B & 1.020--1.040 & 1848 & 0 & 1848 & 55.38\% & GRB080714086--GRB180710062 \\
\code{unic}   & B & 0.253--137.220 & 21 & 0 & 21 & 0.63\% & GRB080906212--GRB170830328 \\
\midrule
Total & A+B & --- & 3337 & 462 & 2875 & 100.00\% & --- \\
\bottomrule
\end{tabular}
\end{table*}

\subsection{Consequences for physical interpretation}\label{subsec:group_implications}
The catalogue period, observation date, reported fit-statistic fields, spectral-catalogue version, and PFLX integration duration characterize measurement or processing conditions rather than intrinsic GRB properties. We use these quantities to define valid comparison groups and to examine systematic effects, but not as physical population variables, because multivariate GRB classifications can change when brightness, fluence, or other variables affected by selection effects are included \citep{chat17,chat18,Toth2019BATSEGMM,hoi19ApS}. PFLX quantities are compared only within the same integration scale. Short--long physical comparisons are based on FLNC quantities, whose time-integrated definition is common to the short and long samples. The A and B groups are used where relevant as separate samples to test whether the results are stable across the catalogue-processing transition.

\section{COMP parameter comparisons across the catalogue periods}\label{sec:compelemzes}

\subsection{Comparison method and effect size}\label{subsec:compstrategy}

For fixed-scale A--B comparisons, we analyse ten COMP quantities: PFLX and FLNC amplitudes, $E_{\rm peak}$ values and indices, PFLX photon and energy fluxes, and FLNC photon and energy fluences. In the multivariate tests, scale quantities spanning large dynamical ranges are log transformed, while the spectral indices remain on their original scale. Short--long comparisons use FLNC quantities only: amplitude, $E_{\rm peak}$, index, photon fluence, and energy fluence.

We summarized the quality of a COMP solution by
\begin{equation}
\left|\frac{\sigma_A}{A}\right|\leq0.4,\qquad
\left|\frac{\sigma_{E_{\rm peak}}}{E_{\rm peak}}\right|\leq0.4,\qquad
\sigma_{\rm index}\leq0.4.
\label{eq:compquality}
\end{equation}

We described pairwise differences by the rank-biserial effect
\begin{equation}
 r_{\rm rb}=P(X_1>X_2)-P(X_2>X_1).
\end{equation}
A positive value means that the first group tends to have larger values. If ties are negligible, $r_{\rm rb}=0.25$ means that a randomly selected value from the first group exceeds one from the second group in approximately 62.5\% of cross-group pairs. We calculated the 95\% intervals by non-parametric bootstrap resampling of GRBs with replacement separately within the two groups being compared. For each fixed-scale A--B comparison, the ten physical Mann--Whitney $p$-values were corrected using the Benjamini--Hochberg procedure \citep{BenjaminiHochberg1995FDR}.

\subsection{Physical COMP parameters across A and B}\label{subsec:compAB}

\begin{figure*}[htbp]
\centering
\includegraphics[width=0.49\textwidth]{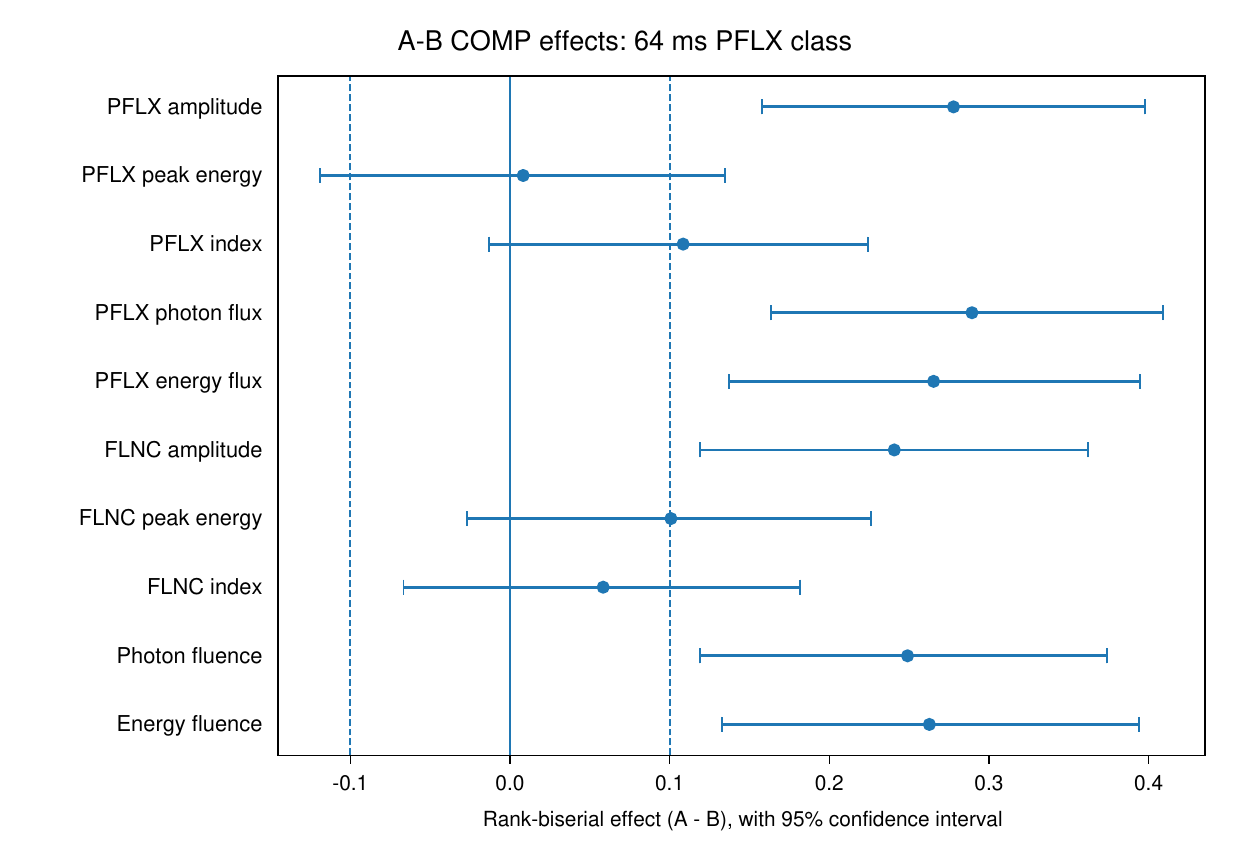}\hfill
\includegraphics[width=0.49\textwidth]{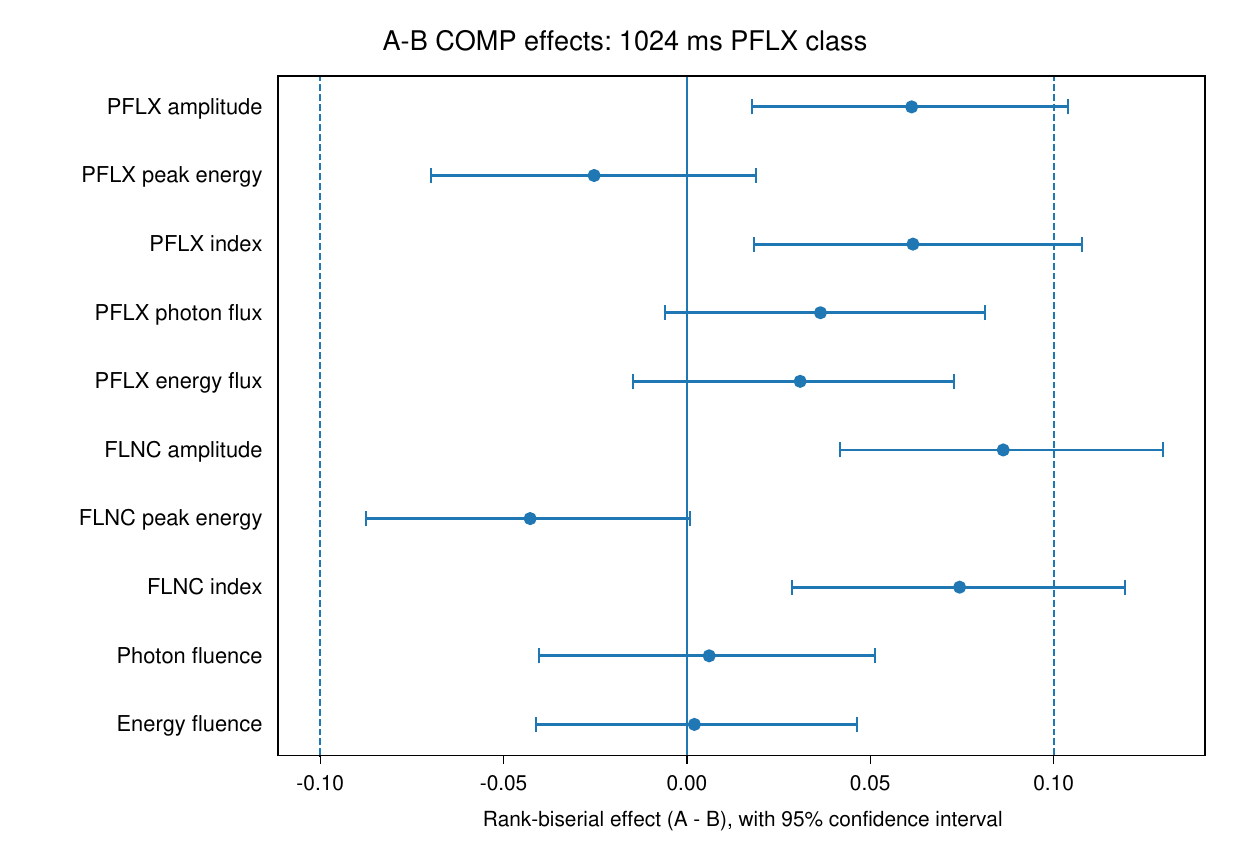}
\caption{Rank-biserial A--B physical effects with 95\% bootstrap intervals within the 64 ms class (left) and the nominal 1024 ms class (right). The dashed vertical lines mark the pre-specified $\pm0.10$ practical-effect margin. Each panel is a fixed-PFLX-scale comparison.}
\label{fig:ABeffects}
\end{figure*}

The shortA and shortB samples contain 103 and 375 GRBs and use the same standard 64 ms PFLX scale. The ten physical effects and the two reported fit-statistic contrasts are given in Table~\ref{tab:compABshort}; the physical effects are shown in the left panel of Fig.~\ref{fig:ABeffects}. The four $E_{\rm peak}$ and index effects are small, with $|r_{\rm rb}|\leq0.108$, and none remains significant after multiplicity correction. On the other hand, all six amplitude, flux, and fluence quantities are higher in shortA, with $r_{\rm rb}=0.241$--0.289; all six remain significant after Benjamini--Hochberg correction (adjusted $p\leq3.1\times10^{-4}$). This brightness difference between A and B is statistically significant in the present 3337-GRB sample. In the auxiliary $T_{90}<2$ s event-level sample, the median \code{FLUX\_64} is 7.72 in A ($N=254$) and 7.44 in B ($N=395$), with no significant A--B difference (Mann--Whitney $p=0.806$). Restricting the same event-level sample to events with a 64 ms, positive-amplitude COMP PFLX fit gives $N=159$ in A and $N=362$ in B, with corresponding \code{FLUX\_64} medians of 9.68 and 7.74. The final 64 ms comparison sample has \code{FLUX\_64} medians of 12.50 in shortA ($N=103$) and 7.70 in shortB ($N=375$); note that shortB also includes 16 bursts with $T_{90}\geq2$ s. These results suggest that sample selection explains part of the A--B brightness difference. We cannot determine exactly why the missing bursts were excluded. We do not interpret this difference physically, and it should be tested in future samples.

\begin{table*}[htbp]
\centering
\scriptsize
\caption{COMP medians and rank-biserial A--B effects within the standard 64 ms PFLX class. Positive effects mean that shortA tends to have larger values. The first ten rows are physical quantities; the final two rows are the reported PFLX and FLNC reduced-statistic contrasts. Amplitude is in photons cm$^{-2}$ s$^{-1}$ keV$^{-1}$, photon flux in photons cm$^{-2}$ s$^{-1}$, energy flux in erg cm$^{-2}$ s$^{-1}$, photon fluence in photons cm$^{-2}$, and energy fluence in erg cm$^{-2}$.}
\label{tab:compABshort}
\begin{tabular}{lrrr}
\toprule
Quantity & shortA median & shortB median & $r_{\rm rb}$ \\
\midrule
PFLX amplitude & 0.0598 & 0.0378 & 0.278 \\
PFLX $E_{\rm peak}$ (keV) & 555 & 532 & 0.008 \\
PFLX index & -0.320 & -0.396 & 0.108 \\
PFLX photon flux & 11.7 & 7.99 & 0.289 \\
PFLX energy flux & $3.78\times10^{-6}$ & $2.34\times10^{-6}$ & 0.265 \\
FLNC amplitude & 0.0362 & 0.0260 & 0.241 \\
FLNC $E_{\rm peak}$ (keV) & 646 & 532 & 0.101 \\
FLNC index & -0.575 & -0.597 & 0.058 \\
Photon fluence & 2.81 & 1.50 & 0.249 \\
Energy fluence & $8.53\times10^{-7}$ & $4.05\times10^{-7}$ & 0.263 \\
PFLX reported reduced-statistic field & 0.466 & 1.019 & -1.000 \\
FLNC reported reduced-statistic field & 0.494 & 1.025 & -1.000 \\
\bottomrule
\end{tabular}
\end{table*}

The longA and longB samples contain 990 and 1848 GRBs and use the same nominal 1024 ms PFLX scale. Table~\ref{tab:compABlong} gives the corresponding effects, while the right panel of Fig.~\ref{fig:ABeffects} shows the physical comparison. All ten physical effects are below 0.10 in absolute value, and six have 95\% bootstrap intervals entirely within the pre-specified $-0.10<r_{\rm rb}<0.10$ margin. Several small effects are statistically detectable because the samples are large, but their magnitudes remain small. Hence, the central COMP spectral parameters and brightness quantities are very similar across A and B in the long sample.

\begin{table*}[htbp]
\centering
\scriptsize
\caption{COMP medians and rank-biserial A--B effects within the nominal 1024 ms PFLX class. Positive effects mean that longA tends to have larger values. The first ten rows are physical quantities; the final two rows are the reported PFLX and FLNC reduced-statistic contrasts. Amplitude is in photons cm$^{-2}$ s$^{-1}$ keV$^{-1}$, photon flux in photons cm$^{-2}$ s$^{-1}$, energy flux in erg cm$^{-2}$ s$^{-1}$, photon fluence in photons cm$^{-2}$, and energy fluence in erg cm$^{-2}$.}
\label{tab:compABlong}
\begin{tabular}{lrrr}
\toprule
Quantity & longA median & longB median & $r_{\rm rb}$ \\
\midrule
PFLX amplitude & 0.0249 & 0.0215 & 0.061 \\
PFLX $E_{\rm peak}$ (keV) & 211.5 & 222.0 & -0.025 \\
PFLX index & -0.753 & -0.802 & 0.062 \\
PFLX photon flux & 3.965 & 3.815 & 0.036 \\
PFLX energy flux & $5.355\times10^{-7}$ & $5.340\times10^{-7}$ & 0.031 \\
FLNC amplitude & 0.009805 & 0.008825 & 0.086 \\
FLNC $E_{\rm peak}$ (keV) & 186 & 201 & -0.043 \\
FLNC index & -0.958 & -1.010 & 0.074 \\
Photon fluence & 36.75 & 35.60 & 0.006 \\
Energy fluence & $4.05\times10^{-6}$ & $3.94\times10^{-6}$ & 0.002 \\
PFLX reported reduced-statistic field & 0.543 & 1.023 & -0.982 \\
FLNC reported reduced-statistic field & 0.5885 & 1.073 & -0.873 \\
\bottomrule
\end{tabular}
\end{table*}

\subsection{Reported fit statistics and catalogue model selection}\label{subsec:modelselection}

The reported fit statistics show much stronger differences than the central COMP spectral parameters. In the 64 ms comparison, every shortA reduced-statistic value lies below the corresponding shortB range in both spectral blocks, giving $r_{\rm rb}=-1.000$. The corresponding effects in the nominal 1024 ms comparison are $-0.982$ for PFLX and $-0.873$ for FLNC. For COMP, the \code{redchisq} and \code{redfitstat} fields are identical in all A rows; in B, equality occurs in only one shortB row in each block, three longB PFLX rows, and no longB FLNC row.

Table~\ref{tab:compmodelquality4} summarizes the associated COMP-best and quality fractions. PFLX fractions are compared only within shortA--shortB and longA--longB, while FLNC uses the common time-integrated definition. Since the fit-statistic and model-selection changes coincide with the statistic-label and catalogue-version transition identified in Sect.~\ref{subsec:catalogtransition}, we treat them as catalogue-processing effects rather than changes in the GRB population.

\begin{table*}[htbp]
\centering
\scriptsize
\caption{COMP model-selection and parameter-quality fractions. ``Both'' means that COMP is the catalogue best model in both PFLX and FLNC. ``Good'' means that all three conditions in Eq.~(\ref{eq:compquality}) are satisfied. PFLX fractions are interpreted only in fixed-scale A--B pairs.}
\label{tab:compmodelquality4}
\begin{tabular}{lrrrrr}
\toprule
Group & PFLX COMP best & FLNC COMP best & Both COMP best & Good PFLX COMP & Good FLNC COMP \\
\midrule
shortA & 75.7\% & 81.6\% & 71.8\% & 83.5\% & 88.3\% \\
shortB & 38.7\% & 59.5\% & 35.7\% & 40.8\% & 63.7\% \\
longA  & 59.5\% & 74.6\% & 53.0\% & 86.4\% & 94.3\% \\
longB  & 50.8\% & 63.4\% & 37.9\% & 62.2\% & 86.0\% \\
\bottomrule
\end{tabular}
\end{table*}

\subsection{Short--long consistency comparison}\label{subsec:flncconsistency}

We also compared short and long GRBs using FLNC quantities only. Table~\ref{tab:flncshortlong} shows that in both A and B, short GRBs have higher $E_{\rm peak}$, less negative COMP indices, and lower photon and energy fluences. We therefore reproduce the well-established short-hard and long-soft spectral difference on both sides of the catalogue transition \citep{kou93,gold12,gru14,spectral_study2,Poolakkil2021GBM10yr,hoi19ApS}. Because this result is already well known, we do not analyse it further here.

\begin{table*}[htbp]
\centering
\scriptsize
\caption{Valid short--long COMP comparison using FLNC variables only. Positive $r_{\rm rb}$ means that the short value tends to be larger. The 95\% confidence interval (CI) is shown in brackets. Amplitude is in photons cm$^{-2}$ s$^{-1}$ keV$^{-1}$, photon fluence in photons cm$^{-2}$, and energy fluence in erg cm$^{-2}$; $E_{\rm peak}$ is in keV and the index is dimensionless.}
\label{tab:flncshortlong}
\begin{tabular}{lrrlrrl}
\toprule
& \multicolumn{3}{c}{A} & \multicolumn{3}{c}{B} \\
\cmidrule{2-4}\cmidrule{5-7}
Quantity & Short median & Long median & $r_{\rm rb}$ [95\% CI] & Short median & Long median & $r_{\rm rb}$ [95\% CI] \\
\midrule
FLNC amplitude & 0.0362 & 0.009805 & 0.714 [0.653, 0.772] & 0.0260 & 0.008825 & 0.642 [0.603, 0.681] \\
FLNC $E_{\rm peak}$ (keV) & 646 & 186 & 0.572 [0.468, 0.665] & 532 & 201 & 0.502 [0.450, 0.553] \\
FLNC index & -0.575 & -0.958 & 0.576 [0.493, 0.657] & -0.597 & -1.010 & 0.577 [0.526, 0.624] \\
Photon fluence & 2.81 & 36.75 & -0.838 [-0.887, -0.782] & 1.50 & 35.60 & -0.918 [-0.939, -0.894] \\
Energy fluence & $8.53\times10^{-7}$ & $4.05\times10^{-6}$ & -0.624 [-0.700, -0.543] & $4.05\times10^{-7}$ & $3.94\times10^{-6}$ & -0.809 [-0.842, -0.773] \\
\bottomrule
\end{tabular}
\end{table*}

The main COMP comparison gives no evidence for either a new short--long division or a physical change at the 2018 boundary. The central COMP spectral-shape parameters are nearly unchanged between A and B, particularly in the long sample. In the short sample, amplitudes, fluxes, and fluences differ significantly, but part of this difference is related to sample selection. We do not know its full origin and cannot conclude that there was a physical change. The much larger changes concern the reported fit statistics and catalogue model-selection fractions. In later population analyses, A and B serve as separate stability samples rather than as possible GRB populations.

\section{Brightness, parameter precision, and the reported COMP statistic}\label{sec:compfitquality}

\subsection{Analysis within B}\label{subsec:compfit_fields}

The statistic labels and the numerical behaviour of the reported fit fields change at the A--B boundary, as shown in Sects.~\ref{subsec:catalogtransition} and \ref{subsec:modelselection}. For this reason, we restrict the analysis of the reported COMP fit statistic in this section to B. PFLX and FLNC are analysed separately, and the 64 ms and nominal 1024 ms PFLX groups are never combined. The detailed comparison of the reported COMP fit fields across A and B is given in Sect.~\ref{app:compfitfield}.

For PFLX, brightness is represented by \code{flux64} in shortB and by \code{flux1024} in longB; for FLNC, the catalogue \code{fluence} is used. We converted all continuous variables to standardized ranks and used heteroskedasticity-consistent type 3 (HC3) robust standard errors \citep{MacKinnonWhite1985HC3}. We include the number of degrees of freedom (DOF) in every model. We analyse the reported fit statistic and parameter errors separately.

\subsection{Brightness, parameter precision, and the reported COMP statistic}\label{subsec:compfit_brightness}

The fit-statistic analysis uses COMP solutions satisfying Eq.~(\ref{eq:compquality}). Within each B group and spectral block, we model the reported statistic as
\begin{equation}
 Z_R(r_{\rm stat}) = b_{\rm br} Z_R(\log_{10}F)+b_{\nu}Z_R(\nu)+\epsilon,
\label{eq:brightness_redchi}
\end{equation}
where $Z_R$ denotes a standardized rank, $F$ is the brightness measure defined above, $r_{\rm stat}$ is the reported COMP reduced-statistic field, and $\nu$ is the COMP number of degrees of freedom.

\begin{figure}[htbp]
\centering
\includegraphics[width=0.90\columnwidth]{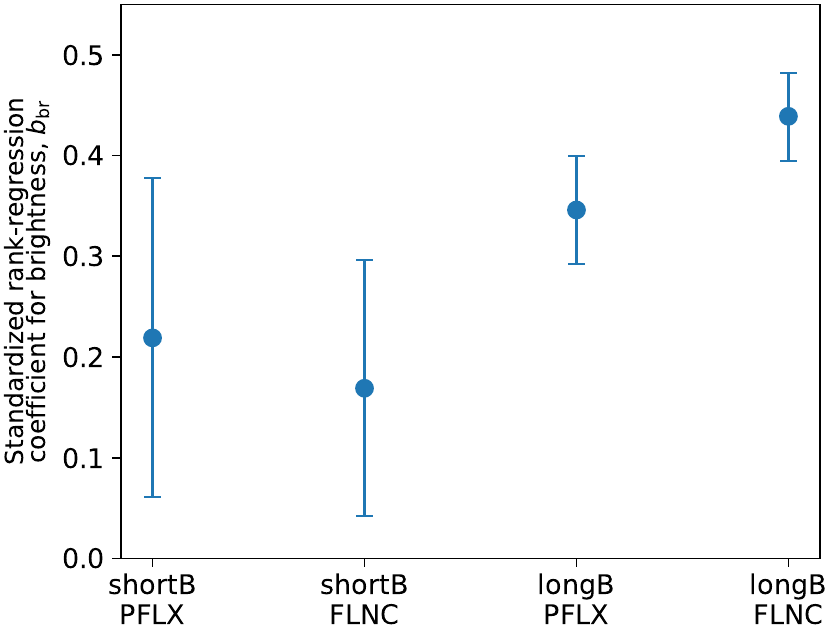}
\caption{Standardized rank-regression coefficient for brightness in the reported COMP reduced-statistic model after controlling for DOF, using COMP solutions that satisfy Eq.~(\ref{eq:compquality}) in B. Error bars are 95\% HC3 confidence intervals.}
\label{fig:brightness_precision}
\end{figure}

We combined the standardized ranks of the amplitude relative error, the $E_{\rm peak}$ relative error, and the absolute index error into the following measure of parameter uncertainty:
\begin{equation}
 U=\frac{1}{3}\left[Z_R\left(\left|\frac{\sigma_A}{A}\right|\right)+
 Z_R\left(\left|\frac{\sigma_{E_{\rm peak}}}{E_{\rm peak}}\right|\right)+
 Z_R(\sigma_{\rm index})\right].
\label{eq:uncertainty_composite}
\end{equation}
To avoid defining the outcome with the same quality threshold used for sample selection, this regression uses all finite rows with valid errors:
\begin{equation}
 U=\gamma_{\rm br} Z_R(\log_{10}F)+\gamma_{\rm stat} Z_R(r_{\rm stat})+
 \gamma_{\nu}Z_R(\nu)+\epsilon.
\label{eq:uncertainty_model}
\end{equation}
Tables~\ref{tab:brightness_redchi} and \ref{tab:precision_model} report the results; $q_{\rm BH}$ denotes the Benjamini--Hochberg-adjusted $p$-value, with values below $10^{-15}$ reported as $q_{\rm BH}<10^{-15}$. Figure~\ref{fig:brightness_precision} illustrates the brightness coefficients.

\begin{table*}[htbp]
\centering
\scriptsize
\caption{Brightness dependence of the reported COMP reduced-statistic field in B. The standardized rank-regression coefficient $b_{\rm br}$ is the brightness term in Eq.~(\ref{eq:brightness_redchi}); brackets give 95\% HC3 confidence intervals. $N_{\rm good}$ is the number of COMP fits satisfying Eq.~(\ref{eq:compquality}).}
\label{tab:brightness_redchi}
\begin{tabular}{llrrrr}
\toprule
Group & Spectrum & $N_{\rm good}$ & $b_{\rm br}$ & 95\% CI & $q_{\rm BH}$ \\
\midrule
shortB & PFLX & 153  & 0.219 & $\left[0.061,0.378\right]$ & 0.0084 \\
shortB & FLNC & 239  & 0.169 & $\left[0.042,0.296\right]$ & 0.0114 \\
longB  & PFLX & 1149 & 0.346 & $\left[0.292,0.400\right]$ & $<10^{-15}$ \\
longB  & FLNC & 1589 & 0.439 & $\left[0.395,0.482\right]$ & $<10^{-15}$ \\
\bottomrule
\end{tabular}
\end{table*}

\begin{table*}[htbp]
\centering
\scriptsize
\caption{Parameter-uncertainty model in B. The coefficients $\gamma_{\rm br}$ and $\gamma_{\rm stat}$ are the brightness and reported reduced-statistic terms in Eq.~(\ref{eq:uncertainty_model}). Brackets give 95\% HC3 confidence intervals. $N_U$ is the number of rows with finite model variables and valid parameter errors used in this regression.}
\label{tab:precision_model}
\begin{tabular}{llrrlrlrl}
\toprule
Group & Spectrum & $N_U$ & $\gamma_{\rm br}$ & 95\% CI & $q_{\rm BH}$ & $\gamma_{\rm stat}$ & 95\% CI & $q_{\rm BH}$ \\
\midrule
shortB & PFLX & 374  & -0.667 & $\left[-0.711,-0.622\right]$ & $<10^{-15}$ & -0.003 & $\left[-0.051,0.046\right]$ & 0.929 \\
shortB & FLNC & 375  & -0.669 & $\left[-0.719,-0.619\right]$ & $<10^{-15}$ & -0.010 & $\left[-0.068,0.047\right]$ & 0.778 \\
longB  & PFLX & 1844 & -0.746 & $\left[-0.765,-0.726\right]$ & $<10^{-15}$ & -0.030 & $\left[-0.051,-0.010\right]$ & 0.0055 \\
longB  & FLNC & 1842 & -0.750 & $\left[-0.769,-0.730\right]$ & $<10^{-15}$ & -0.094 & $\left[-0.115,-0.073\right]$ & $<10^{-15}$ \\
\bottomrule
\end{tabular}
\end{table*}

All four brightness coefficients in the fit-statistic model are positive and remain significant after multiplicity correction, with $b_{\rm br}=0.169$--0.439. Within the same B duration and spectral block, brighter GRBs tend to have larger reported COMP fit statistics. The brightness coefficients in the uncertainty model are strongly negative, $\gamma_{\rm br}=-0.667$ to $-0.750$, so brighter GRBs have smaller formal parameter errors. The independent reported reduced-statistic term is consistent with zero in both shortB analyses and is small and negative in longB. A larger reported COMP fit statistic therefore does not imply a larger formal parameter uncertainty.

\subsection{Evidence for additional spectral curvature}\label{subsec:compfit_alternatives}

To test whether stronger evidence for a more flexible spectral shape accompanies the brightness dependence, we compare COMP and BAND for the same spectrum. We define
\begin{equation}
 \dscb=S_{\rm COMP}-S_{\rm BAND}
 =r_{\rm COMP}\nu_{\rm COMP}-r_{\rm BAND}\nu_{\rm BAND}.
\label{eq:deltaS_CB_compchapter}
\end{equation}
The ten-year catalogue favours BAND over COMP when BAND satisfies its quality criteria and $\dscb>11.83$ \citep{Poolakkil2021GBM10yr}. After allowing for DOF, brightness is positively associated with $\dscb$ in all four B comparisons, with standardized rank-regression coefficients of 0.334--0.483, and BAND is selected more often in the highest brightness quintile. The numerical value of $\beta$ is not required for this comparison. Section~\ref{sec:band} gives the detailed BAND results.

\subsection{Interpretation}\label{subsec:compfit_robustness}

For both PFLX and FLNC in shortB and longB, brighter GRBs have smaller formal COMP parameter errors, but larger reported COMP statistics and COMP--BAND improvements. A spectrum can therefore determine the COMP parameters accurately and still favour a more flexible empirical form. We treat \emph{parameter precision and evidence for departures from the COMP shape as separate properties}.

More photons improve the precision of the fitted parameters and can also reveal small departures from the COMP form. The catalogue alone does not show whether these departures are intrinsic or arise from background treatment, detector response, cross-calibration, or a combination of these effects. Sections~\ref{app:compfitfield} and \ref{app:comprobustness} compare the reported fit fields and repeat the physical tests after extreme reduced-statistic values are removed; no valid physical effect changes by 0.10 or more.

\section{The role and limitations of the PLAW model}\label{sec:plaw}

The simple power law (PLAW) is the simplest of the four catalogue functions. Its amplitude and single photon index are available for almost every burst, and the index gives a direct broad-band measure of spectral hardness. However, PLAW cannot represent curvature or a spectral peak within the GBM energy range \citep{band,2006ApJS..166..298K,gold12,gru14,Goldstein2013BATSE5B}. Although PLAW is most often the poorest numerical fit when all models are compared, it can still be an appropriate catalogue model for some spectra. We examine why this occurs and what independent information remains in its index.

The comparison rules of Sect.~\ref{subsec:pflxscale} are used throughout. PFLX quantities are compared only within the 64 ms or nominal 1024 ms groups, while short--long physical comparisons and duration classification use FLNC quantities only. We assess curvature by comparing PLAW with COMP; SBPL is used only in the additional raw-ranking comparisons.

\subsection{When is PLAW sufficient?}\label{subsec:plaw_strategy}

The ten-year GBM spectral catalogue defines GOOD models from parameter-quality criteria and selects the BEST model by a stepwise fit-statistic comparison \citep{Poolakkil2021GBM10yr}. In our analysis, PLAW requires an absolute amplitude relative error no larger than 0.4 and an index error no larger than 0.4, while COMP additionally requires an absolute $E_{\rm peak}$ relative error no larger than 0.4. The catalogue \code{bestfittingmodel} label remains the primary model-selection result. A regular anomaly in the A-period PLAW amplitude-error field and the catalogue comparison are documented in Sects.~\ref{app:plawamperror} and \ref{app:plawvalidation}.

Following \citet{Poolakkil2021GBM10yr}, we calculate the total reported statistic as
\begin{equation}
 S_m=r_{m,\mathrm{fit}}\,\nu_m,
\end{equation}
Here $r_{m,\mathrm{fit}}$ is the model-specific catalogue \code{redfitstat} value and $\nu_m$ is its number of degrees of freedom; \code{redchisq} is not used in this calculation.
We define the PLAW--COMP curvature statistic
\begin{equation}
 \Delta S_{\rm PC}=S_{\rm PLAW}-S_{\rm COMP}.
\label{eq:plaw_deltaS}
\end{equation}
The catalogue selects COMP over PLAW when the COMP solution is GOOD and $\Delta S_{\rm PC}>8.58$. Since the reported statistic changes at the catalogue boundary, we apply this threshold directly only in B; A is used mainly as a comparison with the catalogue label.

Table~\ref{tab:plaw_sufficiency} summarizes the B results. In the subsample in which all four models are well measured, PLAW has the largest raw statistic in 94.7\% of shortB FLNC and 99.6\% of longB FLNC spectra. Nevertheless, PLAW can remain catalogue BEST when COMP is not well measured or when the improvement does not exceed the catalogue threshold. The calculated PLAW--COMP rule agrees with the catalogue result in 99.2--99.6\% of the B tests.

\begin{table*}[htbp]
\centering
\scriptsize
\caption{Compact PLAW sufficiency comparison in B. ``Raw worst'' is the fraction for which PLAW has the largest reported reduced-statistic value among the four models. ``All four good'' gives the number of events followed in parentheses by the PLAW-raw-worst fraction. ``PLAW+COMP good'' gives the number of events followed by the fraction with $\Delta S_{\rm PC}>8.58$. Agreement compares this calculated rule with the catalogue non-PLAW result.}
\label{tab:plaw_sufficiency}
\resizebox{\textwidth}{!}{%
\begin{tabular}{llrrrrr}
\toprule
Group & Spectrum & Raw worst (\%) & PLAW best (\%) & \shortstack{All four good\\$N$ (raw worst \%)} & \shortstack{PLAW+COMP good\\$N$ ($\Delta S_{\rm PC}>8.58$ \%)} & Agreement (\%) \\
\midrule
shortB & PFLX & 47.2 & 60.3 & 5 (80.0\%)   & 153 (98.0\%)  & 99.3 \\
shortB & FLNC & 67.2 & 38.1 & 19 (94.7\%)  & 239 (97.5\%)  & 99.6 \\
longB  & PFLX & 61.1 & 43.0 & 216 (95.4\%) & 1149 (91.0\%) & 99.6 \\
longB  & FLNC & 84.4 & 18.1 & 489 (99.6\%) & 1589 (93.7\%) & 99.2 \\
\bottomrule
\end{tabular}%
}
\end{table*}

Thus, for faint spectra PLAW can be an appropriate simple catalogue choice because curvature is not always measured reliably or the improvement over PLAW is too small. This does not imply that the underlying spectrum is necessarily an unbroken power law.

\subsection{Brightness and interpretation of the PLAW index}\label{subsec:plaw_brightness}

Within B, we used rank-based regressions controlling for DOF and found a positive association of brightness with both the reported PLAW fit statistic and the curvature statistic $\Delta S_{\rm PC}$. The standardized rank-regression coefficients are 0.368--0.763 for the reported PLAW statistic and 0.776--0.844 for $\Delta S_{\rm PC}$. The catalogue PLAW-best fraction, however, falls sharply toward higher brightness, while well-measured COMP solutions become much more frequent. We find that PLAW is selected often for faint spectra because curvature is difficult to establish there, whereas brighter spectra increasingly require a curved model.

\begin{figure}[htbp]
\centering
\includegraphics[width=0.90\columnwidth]{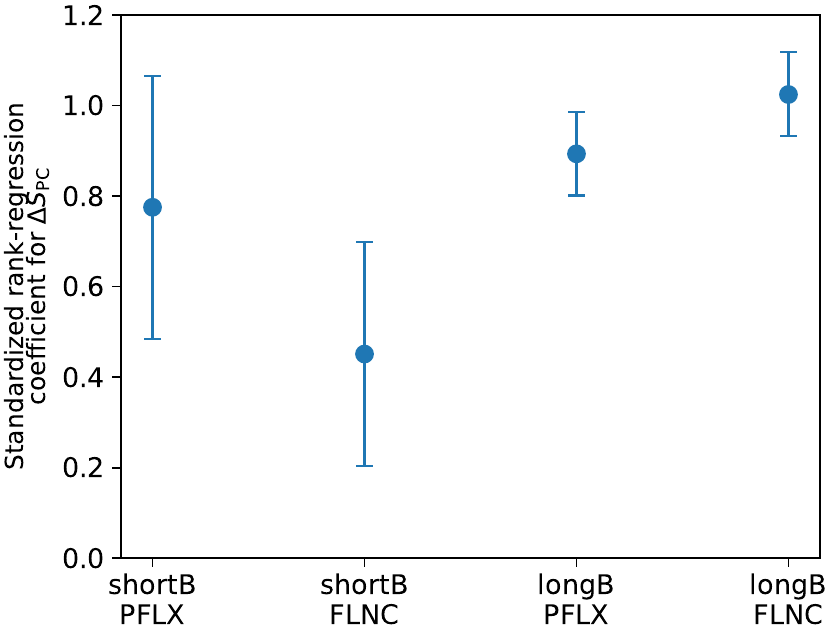}
\caption{Growth of the absolute PLAW--COMP index difference with $\Delta S_{\rm PC}$. Points show standardized rank-regression coefficients after controlling for brightness and DOF; error bars are 95\% HC3 confidence intervals.}
\label{fig:plaw_precision_bias}
\end{figure}

The PLAW index error decreases with brightness, with simultaneous brightness effects of $-0.579$ to $-0.696$. However, a small formal error does not make the PLAW index a physical low-energy slope. For B events with well-measured PLAW and COMP fits, the PLAW index is systematically more negative than the COMP low-energy index by median amounts of 0.580--0.902, and the absolute disagreement grows with $\Delta S_{\rm PC}$. Table~\ref{tab:plaw_comp_divergence} in Sect.~\ref{app:plawmodeldependence} gives the detailed model-dependent differences, and Fig.~\ref{fig:plaw_precision_bias} illustrates this model dependence.

A PLAW index can be statistically precise even when the one-slope model does not describe the full measured spectrum. For curved spectra, it should be interpreted as an effective broad-band slope and not as a quantity equivalent to the COMP or BAND low-energy index.

\subsection{Information carried by the FLNC PLAW index}\label{subsec:plaw_classifier}

We find that the FLNC PLAW index is stable across the catalogue transition: the shortA--shortB rank effect is $-0.001$ and the longA--longB effect is 0.007. In contrast, the FLNC short--long effects are much larger, 0.709 in A and 0.698 in B, consistent with the established short--hard and long--soft difference.

For the full 3337-GRB sample, the FLNC PLAW index gives a receiver operating characteristic (ROC) area under the curve (AUC) of 0.847. The A and B AUCs, 0.854 and 0.846, are nearly identical. For the common sample with a well-measured COMP shape, the PLAW index reaches cross-validated AUC 0.869, compared with 0.886 for COMP index plus $\log E_{\rm peak}$; adding PLAW to the COMP shape leaves the AUC unchanged at 0.886 (Table~\ref{tab:plaw_information}).

\begin{table*}[htbp]
\centering
\scriptsize
\caption{Repeated 10-fold cross-validated duration classifiers used to assess the information in the FLNC PLAW index. The first row uses all 3337 GRBs; the final three use the common FLNC sample with a COMP fit having small parameter errors.}
\label{tab:plaw_information}
\begin{tabular}{llrrr}
\toprule
Model & Sample & $N$ & ROC AUC & Balanced accuracy \\
\midrule
PLAW index & Full catalogue & 3337 & 0.847 & 0.773 \\
PLAW index & Common COMP sample & 2870 & 0.869 & 0.790 \\
COMP index + $\log E_{\rm peak}$ & Common COMP sample & 2870 & 0.886 & 0.818 \\
PLAW + COMP shape & Common COMP sample & 2870 & 0.886 & 0.816 \\
\bottomrule
\end{tabular}
\end{table*}

The PLAW index is available for nearly all bursts and acts as an effective hardness measure. In a sample with a well-measured curved COMP shape, however, it contributes almost no independent duration information. For later population analyses, we use PLAW in a separate variable set or as a robustness test rather than adding it to a COMP representation with small parameter errors. Sections~\ref{app:plawvalidation} and \ref{app:plawrobustness} give the additional catalogue-validation and robustness tests.

\section{BAND low-energy parameters and the limitations of the high-energy index}\label{sec:band}

The BAND function extends COMP with a power-law high-energy branch \citep{band,2006ApJS..166..298K,gold12,gru14}. Its principal shape parameters are the low-energy index $\alpha$, the high-energy index $\beta$, and $E_{\rm peak}$. Here we examine which of these parameters can be used consistently in later population analyses. We use $\alpha$, $E_{\rm peak}$, and the COMP--BAND fit improvement in the main analysis, while $\beta$ is used only to study uncertainty and selection effects.

The comparison rules established in Sect.~\ref{subsec:pflxscale} remain in force. Short and long PFLX quantities are never combined because their spectra were accumulated over 64 ms and nominal 1024 ms, respectively. Thus, short--long comparisons and duration classification are made with FLNC variables only, while A--B PFLX comparisons are restricted to a fixed integration scale.

\subsection{Usable BAND parameters}\label{subsec:band_strategy}

The ten-year catalogue defines GOOD models from parameter errors and selects the BEST model through successive fit-statistic comparisons \citep{Poolakkil2021GBM10yr}. For the main BAND shape analysis we use
\begin{equation}
\begin{aligned}
 \text{$0<$}\sigma_{\alpha}&<0.51, &
 \text{$0<$}\left|\frac{\sigma_{E_{\rm peak}}}{E_{\rm peak}}\right|&<0.43,\\
 E_{\rm peak}&>0, & \text{$-2<\alpha<5$}&.
\end{aligned}
\label{eq:band_core_criterion}
\end{equation}
without requiring a usable $\beta$. A separate high-energy-index condition requires a sufficiently small parameter error.

Table~\ref{tab:band_sample} shows the resulting sample sizes. The $\alpha$--$E_{\rm peak}$ criterion is satisfied by 37.6--94.6\% of the eight principal group--spectrum samples, whereas the usable-$\beta$ fraction is only 1.9--37.5\%. Requiring all BAND parameters therefore removes most events, especially from the short samples.

\begin{table*}[htbp]
\centering
\scriptsize
\caption{Fraction of BAND fits satisfying the $\alpha$--$E_{\rm peak}$ criterion and the high-energy-index condition. Percentages are relative to the full group and spectral block. The full criterion includes the $\beta$ uncertainty and is shown only to quantify the loss of events.}
\label{tab:band_sample}
\begin{tabular}{llrrrrr}
\toprule
Group & Spectrum & $N$ & $\alpha+E_{\rm peak}$ usable & $\beta$ usable & Full criterion & BAND best \\
\midrule
shortA & PFLX & 103  & 81 (78.6\%)   & 2 (1.9\%)   & 0 (0.0\%)   & 0 (0.0\%) \\
shortA & FLNC & 103  & 88 (85.4\%)   & 2 (1.9\%)   & 2 (1.9\%)   & 1 (1.0\%) \\
shortB & PFLX & 375  & 141 (37.6\%)  & 33 (8.8\%)  & 14 (3.7\%)  & 2 (0.5\%) \\
shortB & FLNC & 375  & 224 (59.7\%)  & 52 (13.9\%) & 38 (10.1\%) & 3 (0.8\%) \\
longA  & PFLX & 990  & 866 (87.5\%)  & 55 (5.6\%)  & 47 (4.7\%)  & 6 (0.6\%) \\
longA  & FLNC & 990  & 937 (94.6\%)  & 200 (20.2\%)& 198 (20.0\%)& 28 (2.8\%) \\
longB  & PFLX & 1848 & 1150 (62.2\%) & 444 (24.0\%)& 348 (18.8\%)& 77 (4.2\%) \\
longB  & FLNC & 1848 & 1471 (79.6\%) & 693 (37.5\%)& 605 (32.7\%)& 212 (11.5\%) \\
\bottomrule
\end{tabular}
\end{table*}

For comparison with COMP, we define
\begin{equation}
 \dscb=S_{\rm COMP}-S_{\rm BAND}
 =r_{\rm COMP}\nu_{\rm COMP}-r_{\rm BAND}\nu_{\rm BAND}.
\label{eq:band_deltaS}
\end{equation}
The catalogue favours BAND over COMP when BAND satisfies its quality requirements and $\dscb>11.83$ \citep{Poolakkil2021GBM10yr}. The calculated rule agrees with the catalogue COMP-versus-BAND outcome in 98.8--100\% of the B tests; Appendix~\ref{app:banddetails} gives the detailed validation and the comparison with the published ten-year medians.

The numerical $\beta$ values are much less reliable than $\alpha$ and $E_{\rm peak}$. In B, the median reported $\beta$ error ranges from 0.490 in longB FLNC to 99.9 in shortB PFLX, and even in longB FLNC 42.6\% of the fits have $\sigma_{\beta}>1$. In A, exact $\beta=-2$ occurs in 85.4\% and 86.4\% of shortA PFLX and FLNC fits and in 50.3\% and 31.9\% of longA PFLX and FLNC fits. The detailed error test is given in Appendix~\ref{app:banddetails}.

An exact $\beta=-2$ value does not by itself invalidate the low-energy parameters. In longA FLNC, fits with exact $\beta=-2$ and fits with other $\beta$ values have very similar median $\alpha$ and $E_{\rm peak}$. Therefore, we retain $\alpha$ and $E_{\rm peak}$ when their own errors satisfy Eq.~(\ref{eq:band_core_criterion}), but exclude $\beta$ from the main physical analysis. The full error distributions and sample-reduction tests are reported in Appendix~\ref{app:banddetails}.

\subsection{Brightness, curvature, and model dependence}\label{subsec:band_brightness}

We restrict the absolute fit-statistic analysis to B. For each duration and spectral block, we modelled the rank-normalised BAND statistic and $\dscb$ as functions of brightness while controlling for DOF. Brightness is positively associated with both quantities: the standardized rank-regression coefficients for $\dscb$ are 0.334--0.483, and the catalogue BAND-best fraction rises from zero in every lowest brightness quintile to 2.7--40.8\% in the corresponding highest quintile. Brighter spectra therefore more often require a high-energy shape more flexible than the COMP exponential cutoff.

Brightness also strongly reduces the statistical uncertainty of $\alpha$ and $E_{\rm peak}$: the brightness coefficients for their uncertainty composite are $-0.776$ to $-0.856$. The independent BAND-statistic coefficient is zero or small and negative, so a larger reported fit statistic does not imply less precise low-energy parameters. The fraction of fits with usable $\beta$ also increases with brightness, but less rapidly than the fraction with usable $\alpha$ and $E_{\rm peak}$. The detailed error and brightness-quintile results are given in Appendix~\ref{app:banddetails}.

\begin{figure}[htbp]
\centering
\includegraphics[width=0.90\columnwidth]{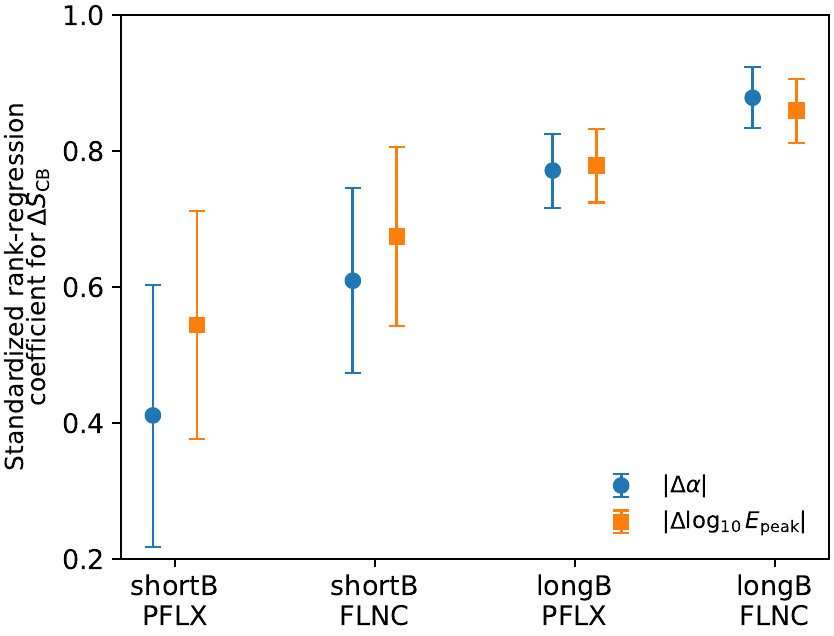}
\caption{Growth of the absolute COMP--BAND parameter differences with $\dscb$. Points show standardized rank-regression coefficients after controlling for brightness and DOF; error bars are 95\% HC3 confidence intervals.}
\label{fig:band_precision_divergence}
\end{figure}

Larger $\dscb$ values are also associated with larger differences between the COMP and BAND parameters fitted to the same spectrum. After allowing for brightness and DOF, we find effects of 0.411--0.878 on $|\Delta\alpha|$ and 0.544--0.859 on $|\Delta\log_{10}E_{\rm peak}|$. Figure~\ref{fig:band_precision_divergence} illustrates this model dependence.

A small formal error does not guarantee that the inferred low-energy index or peak energy is independent of the selected empirical model. Better signal quality can reveal the need for an additional high-energy branch before the numerical slope of that branch is measured accurately.

\subsection{Information carried by $\alpha$ and $E_{\rm peak}$}\label{subsec:band_classifier}

The A--B spectral-shape effects are small compared with the established short--long difference. In the FLNC sample not requiring $\beta$, the A--B effects on $\log E_{\rm peak}$ are $-0.027$ in the 64 ms group and $-0.031$ in the nominal 1024 ms group, while the corresponding $\alpha$ effects are 0.161 and 0.068. The larger short--long effects reproduce the familiar short-hard and long-soft relation and are used only as a consistency comparison. Detailed group effects and robustness results are given in Appendix~\ref{app:banddetails}.

The FLNC $\alpha+\log E_{\rm peak}$ model gives repeated-cross-validation ROC AUC 0.877 and balanced accuracy 0.802 in the 2735-event BAND shape sample. Table~\ref{tab:band_information} gives the more relevant common-sample comparison. The COMP model is slightly better than BAND, while adding the BAND parameters to the COMP shape gives no meaningful improvement.

\begin{table*}[htbp]
\centering
\scriptsize
\caption{Repeated ten-fold cross-validated duration classifiers used to assess the information in the BAND low-energy parameters. The first row uses the full BAND shape sample that does not require $\beta$; the final three rows use the common BAND--COMP sample.}
\label{tab:band_information}
\begin{tabular}{llrrr}
\toprule
Model & Sample & $N$ & ROC AUC & Balanced accuracy \\
\midrule
BAND $\alpha+\log E_{\rm peak}$ & BAND shape sample & 2735 & 0.877 & 0.802 \\
BAND $\alpha+\log E_{\rm peak}$ & Common BAND--COMP sample & 2675 & 0.875 & 0.801 \\
COMP index+$\log E_{\rm peak}$ & Common BAND--COMP sample & 2675 & 0.885 & 0.808 \\
BAND+COMP shape & Common BAND--COMP sample & 2675 & 0.884 & 0.807 \\
\bottomrule
\end{tabular}
\end{table*}

Training on A and testing on B gives AUC 0.870; the reverse transfer gives 0.892. The close values indicate that the BAND low-energy hardness information is stable across the catalogue transition. On the common sample with a well-measured COMP shape, adding BAND gives little independent classification information. Appendix~\ref{app:banddetails} gives the remaining classifier results and A--B comparisons.

\subsection{Interpretation}\label{subsec:band_interpretation}

BAND $\alpha$ and $E_{\rm peak}$ are useful spectral-shape parameters for a large part of the catalogue when their own uncertainties are controlled. The COMP--BAND improvement $\dscb$ is also useful for identifying spectra that require a high-energy form more flexible than COMP.

The numerical $\beta$ value cannot be used reliably as a common physical variable for the full catalogue. Its uncertainty is often large, and exact $\beta=-2$ values are common in A. After the quality cuts, usable $\beta$ values occur mainly among brighter events. In later population analyses, we use the BAND low-energy parameters in a separate variable set. We do not add them to a COMP shape with small parameter errors as independent variables; $\beta$ is used only for individual bursts for which its error is sufficiently small.

\section{Discussion}\label{sec:discussion}
\subsection{Catalogue-processing transition}
The catalogue information indicates a processing change at \code{GRB180712358}. Before the boundary, all eight model- and spectrum-specific statistic fields contain \code{Castor C-STAT}. After the boundary, all non-missing COMP, BAND, and SBPL fields contain \code{PSTAT}, with only five GRBs retaining Castor C-STAT in one or both PLAW fields. At the same row, the spectral-catalogue version changes from \code{scatalog}=3 or 4 to \code{scatalog}=5, and the numerical relation between the reported fit fields also changes. The A--B differences in the central COMP, PLAW, and BAND spectral-shape parameters are much smaller than those in the reported fit and uncertainty fields. In the 64 ms COMP sample, amplitudes, fluxes, and fluences are higher in shortA than in shortB, but part of this difference may be due to sample selection. The nominal 1024 ms physical quantities are very similar across A and B.

Changes in the GBM triggering configuration, detector response, and background conditions may also affect the observed GRB population and the derived spectral parameters. Previous GBM studies document modifications of triggering criteria and describe instrument gain stabilization and variable background contributions \citep{vonKienlin2020GBM4,2009ApJ...702..791M,Biltzinger2020Background}. However, we found no documented sudden instrumental change at the July 2018 boundary. These effects may influence the measured parameters. We cannot show that they caused the change in the reported fit statistics.

\subsection{Measurement scale and short--long inference}
The standard PFLX spectra for short and long bursts are accumulated over nominal 64 and 1024 ms, respectively \citep{gold12,gru14,Poolakkil2021GBM10yr}. Hence, PFLX values can be compared only within the same nominal integration-time class. Combining short and long PFLX values would mix spectral differences with the different integration windows and with the known energy and detector dependence of duration measurements \citep{kos96,band03,qin13}. Because FLNC has the same time-integrated definition for both groups, it provides the better basis for short--long comparisons. In all three spectral-model chapters, the FLNC comparison is used only as a consistency comparison and recovers the familiar short--hard and long--soft structure \citep{kou93,gold12,Poolakkil2021GBM10yr} without the PFLX-scale ambiguity.

\subsection{Brightness differences between A and B and within B}
The 64 ms A--B comparison and the within-B analysis show different brightness effects. ShortA has higher amplitudes, fluxes, and fluences than shortB. Within B, brighter bursts have smaller parameter errors and larger reported fit statistics. More photons can reduce statistical errors while making small departures from a model easier to detect. The reported COMP and BAND statistics, the PLAW--COMP and COMP--BAND improvements, and the frequency of curved-model selection all increase with brightness. Stronger curvature evidence is also associated with larger differences between parameters fitted by different empirical models. Thus, small formal errors do not imply model-independent parameters. These results agree with bright-burst BATSE spectroscopy and the successive GBM spectral catalogues \citep{2006ApJS..166..298K,gold12,gru14,yu_evolution,Poolakkil2021GBM10yr}.

\subsection{What the empirical models measure}
PLAW is often the poorest numerical fit when all models are well measured, but it is an appropriate simple catalogue choice for many faint spectra. Its index is stable and useful as an effective broad-band hardness measure. COMP gives a low-energy slope and peak energy for a large sample, but for bright spectra its exponential high-energy cutoff is more often found to be inadequate. BAND supplies a more flexible high-energy branch; its $\alpha$ and $E_{\rm peak}$ are useful for many bursts, while $\beta$ is frequently too uncertain for physical interpretation in the full catalogue.

PLAW, COMP, and BAND contain partly overlapping low-energy information. The PLAW index is available for almost all bursts and measures effective hardness, whereas COMP and BAND give model-specific low-energy indices and peak energies. In the common samples, duration classification is not improved by adding PLAW or BAND to a well-measured COMP shape. We analyse these parameter sets separately rather than treating them as independent spectral variables.

A well-measured BAND $\alpha$ need not be a unique asymptotic slope. An additional low-energy break can be hidden by low signal-to-noise ratio, detector bandpass, or proximity to $E_{\rm peak}$; the fitted index may then average unresolved spectral segments \citep{Toffano2021LowEnergySlope}. When well measured, the SBPL low-energy index $\lambda_1$ follows the same broad hardness direction, but it is available only for a brightness-biased subset and cannot serve as a common parameter for the full catalogue. The high-energy index $\lambda_2$ is not measured with comparable accuracy in all groups. This behaviour agrees with the empirical-model ordering and parameter distributions in the BATSE and GBM spectral catalogues \citep{band,smothly,Kumar16,2006ApJS..166..298K,gold12,gru14,2018Rac,Poolakkil2021GBM10yr}. A catalogue model label alone is therefore insufficient to identify a unique emission mechanism, although spectral groups found in population studies can be compared with synchrotron and photospheric expectations \citep{Kumar5,evol_2,acuneryde18,2015AdAst2015E..22P}. Detailed analyses of individual bright bursts sometimes require additional thermal or high-energy components beyond the four standard catalogue functions, and physical synchrotron fits need not correspond uniquely to the empirical BAND parameters \citep{Kumar18,evol_5,thermal1,peter_meszaros_2014,Burgess2014Synchrotron}. Catalogue labels are empirical descriptions of the spectra, not complete radiation models.

\subsection{Implications for population analyses}\label{subsec:population_implications}
For full-catalogue population analyses, we keep spectral-shape information separate from brightness, fit-statistic behaviour, catalogue version, and fit-statistic fields. FLNC has a common spectral definition for short and long bursts, whereas PFLX quantities are comparable only at a fixed integration scale. Low-energy indices and peak energies are used as spectral parameters when their errors are sufficiently small. Fit statistics, parameter errors, catalogue version, and PFLX integration duration are used instead to study measurement and selection effects, not to define astrophysical populations. This distinction is important because the number and interpretation of GRB groups depend on the variables and statistical method used \citep{hor98,muk98,chat17,chat18,acuneryde18,Toth2019BATSEGMM,Jespersen2020,Salmon2022FirstSecond,Steinhardt2023,Chen2024FermiSpectralML,Zhu2024,Mehta2024GRBDiversity,Dimple2024FermiGBMML,Chen2025DeepLearning,Modak2025BinaryClustering,RafteryDean2006VariableSelection}. In later analyses, A and B serve as separate stability samples rather than as possible population groups. The model-based variable sets are also compared separately instead of being combined as independent inputs.

\subsection{Limitations}
The present study uses Fermi/GBM catalogue data and, unlike the event-level and time-resolved spectral studies from which the catalogue products are constructed, does not reanalyse counts from Time-Tagged Event (TTE) or Continuous Spectroscopy (CSPEC) data with detector response matrices \citep{gold12,gru14,yu_evolution,Poolakkil2021GBM10yr}. The catalogue does not give all the details of the software, energy channels, background treatment, and stored fit values. We therefore cannot calibrate the reported statistics in A and B on the same basis. The published definitions show possible reasons for the difference, but not how much each contributes. To test this, the same count spectra and response matrices would need to be fitted with both methods. The catalogue does not establish the origin of the shortA--shortB brightness difference, and this result should be tested with future data. A lack of improvement in supervised duration classification does not by itself exclude additional information for unsupervised clustering. Finally, the supervised classifiers are used only as secondary tests of spectral information. They do not establish unsupervised populations or progenitor classes; this is important because the number of GRB clusters depends on the method used \citep{2022MNRAS517.5770Z,Steinhardt2023,Chen2025DeepLearning,Modak2025BinaryClustering}.

\section{Conclusions}\label{sec:conclusions}
The main conclusions are as follows.
\begin{enumerate}
\item The catalogue changes sharply at \code{GRB180712358}. The catalogue version, statistic name, and relation between the fit fields change together. This points to a change in catalogue processing.
\item Standard PFLX spectra for short and long events use different integration durations. PFLX variables must therefore be compared within fixed 64 or nominal 1024 ms groups; short--long physical inference should use FLNC quantities.
\item Across both fixed-scale comparisons, the central COMP spectral-shape parameters are similar in A and B. In the 64 ms sample, amplitudes, fluxes, and fluences are higher in shortA than in shortB; the larger short-burst sample suggests that sample selection may explain part of this difference. We do not know the exact reason for the selection or whether another difference remains. In the nominal 1024 ms sample, all ten physical rank-biserial point estimates satisfy $|r_{\rm rb}|<0.10$, our pre-specified threshold for a practically small A--B difference. The reported fit statistics and model-selection fractions, however, change much more strongly across the catalogue boundary.
\item Within B, a separate brightness dependence is present: brighter bursts have smaller formal parameter errors but larger reported fit statistics and stronger evidence for spectral curvature. Thus, parameter precision and evidence for departures from the fitted spectral form are distinct properties.
\item The FLNC PLAW index is a widely available effective hardness measure, although in curved spectra it is not equivalent to a physical low-energy index and it adds little independent duration information when a COMP shape with small parameter errors is available.
\item BAND $\alpha$ and $E_{\rm peak}$ can be used for many bursts when their own errors are controlled; a well-measured $\beta$ is not required. However, $\beta$ is not reliable across the full catalogue, and larger COMP--BAND fit improvements are associated with larger model-dependent differences in $\alpha$ and $E_{\rm peak}$. For common samples, BAND gives little classification information beyond a well-measured COMP shape.
\item PLAW and BAND parameters should be examined in separate model-based sets and not combined with COMP as independent spectral variables. The SBPL $\lambda_1$ contains useful low-energy information when well measured, but it is available only for a brightness-dependent fraction of the sample, while $\lambda_2$ is not measured with comparable accuracy across the catalogue.
\end{enumerate}
These results provide a basis for later multivariate population analyses without assuming a particular number of GRB groups. Our second paper will investigate the multivariate structure of the Fermi/GBM population.

\hyphenation{Post-Script Sprin-ger}

\begin{appendix}
\nolinenumbers
\section{Detailed reported reduced-statistic model-rank comparisons}\label{app:rankchecks}

The supporting model-rank results used in Sect.~\ref{subsec:modelchecks} are collected here. Tables~\ref{tab:flnc_legroszab_chi2model_3378} and \ref{tab:pflx_legroszab_chi2model_3378} give the distributions of the worst-ranked FLNC and PFLX models, respectively, and Table~\ref{tab:biggest_chi2_flnc_pflx_3378} gives their joint distribution. Table~\ref{tab:fe26rmi3337chi2dif123} gives the reported reduced-statistic separations among COMP, BAND, and SBPL ranks for the 3337-GRB sample. After PLAW is removed from the raw-rank comparison, Table~\ref{tab:noplaw-3x3} gives the FLNC--PFLX cross-tabulation of the curved model with the smallest reported reduced-statistic value.

\begin{table}[ht]
\centering
\caption{Distribution of the worst-ranked fluence-spectrum models for the 3378 GRBs. The ranking is based on the largest reported reduced-statistic value among the four spectral models.}
\label{tab:flnc_legroszab_chi2model_3378}
\begin{tabular}{lr}
\hline
Worst FLNC model & No. GRBs \\
\hline
\texttt{flnc\_plaw} & 2900 \\
\texttt{flnc\_comp} & 246 \\
\texttt{flnc\_band} & 105 \\
\texttt{flnc\_sbpl} & 127 \\
\hline
Total & 3378 \\
\hline
\end{tabular}
\end{table}

\begin{table}[ht]
\centering
\caption{Distribution of the worst-ranked peak-flux-spectrum models for the 3378 GRBs. The ranking is based on the largest reported reduced-statistic value among the four spectral models.}
\label{tab:pflx_legroszab_chi2model_3378}
\begin{tabular}{lr}
\hline
Worst PFLX model & No. GRBs \\
\hline
\texttt{pflx\_plaw} & 2323 \\
\texttt{pflx\_comp} & 536 \\
\texttt{pflx\_band} & 246 \\
\texttt{pflx\_sbpl} & 273 \\
\hline
Total & 3378 \\
\hline
\end{tabular}
\end{table}

\begin{table*}[ht]
\centering
\caption{Cross-tabulation of the worst-ranked FLNC and PFLX spectral models for the 3378 GRBs. On each side, the worst-ranked model is the one with the largest reported reduced-statistic value among the four spectral models.}
\label{tab:biggest_chi2_flnc_pflx_3378}
\begin{tabular}{lrrrrr}
\hline
Worst FLNC $\backslash$ worst PFLX & \texttt{pflx\_plaw} & \texttt{pflx\_comp} & \texttt{pflx\_band} & \texttt{pflx\_sbpl} & Row total \\
\hline
\texttt{flnc\_plaw} & 2241 & 333 & 140 & 186 & 2900 \\
\texttt{flnc\_comp} & 29 & 148 & 41 & 28 & 246 \\
\texttt{flnc\_band} & 22 & 23 & 36 & 24 & 105 \\
\texttt{flnc\_sbpl} & 31 & 32 & 29 & 35 & 127 \\
\hline
Column total & 2323 & 536 & 246 & 273 & 3378 \\
\hline
\end{tabular}
\end{table*}

The distribution below gives the reported reduced-statistic rank separations for the same spectra in the complete 3337-GRB spectral-parameter sample. For each GRB, we ranked the reported reduced-statistic values of COMP, BAND, and SBPL separately for FLNC and PFLX. Rank 1 denotes the largest value, corresponding to the worst numerical fit among the three models; rank 2 is the middle value; and rank 3 denotes the smallest value, corresponding to the best numerical fit. The quantities \code{flnc3chi12}, \code{flnc3chi23}, and \code{flnc3chi13} are the differences between the reported reduced-statistic values at ranks 1 and 2, 2 and 3, and 1 and 3, respectively, for the FLNC fits. The quantities \code{pflx3chi12}, \code{pflx3chi23}, and \code{pflx3chi13} are defined in the same way for the PFLX fits. Table~\ref{tab:fe26rmi3337chi2dif123} gives the numbers of GRBs in each interval of these six differences.

These differences describe the reported catalogue values. The fields and their normalization differ between A and B. Therefore, the same difference does not have the same likelihood-ratio meaning or significance in both periods. We use Table~\ref{tab:fe26rmi3337chi2dif123} only to show the numerical separation.

\begin{table*}[htbp]
\small
\caption{Reported reduced-statistic model-rank comparisons among the COMP, BAND, and SBPL fits for the 3337-GRB spectral-parameter sample. Each of the six variables contains one value per GRB.}
\label{tab:fe26rmi3337chi2dif123}
\begin{tabular}{lrrrrrr}
\toprule
Class & flnc3chi12 & flnc3chi23 & flnc3chi13 & pflx3chi12 & pflx3chi23 & pflx3chi13 \\
\midrule
$\leq 0.001$ & 967 & 1541 & 355 & 1092 & 1506 & 463 \\
$> 0.001$--$\leq 0.003$ & 628 & 851 & 668 & 593 & 871 & 637 \\
$> 0.003$--$\leq 0.005$ & 287 & 320 & 389 & 284 & 326 & 388 \\
$> 0.005$--$\leq 0.007$ & 206 & 131 & 297 & 159 & 156 & 248 \\
$> 0.007$--$\leq 0.01$ & 205 & 132 & 272 & 159 & 140 & 240 \\
$> 0.01$--$\leq 0.02$ & 388 & 171 & 515 & 361 & 154 & 479 \\
$> 0.02$--$\leq 0.03$ & 167 & 74 & 236 & 173 & 68 & 229 \\
$> 0.03$--$\leq 0.04$ & 108 & 28 & 138 & 123 & 38 & 161 \\
$> 0.04$--$\leq 0.05$ & 65 & 20 & 80 & 62 & 18 & 100 \\
$> 0.05$--$\leq 0.07$ & 103 & 18 & 116 & 92 & 13 & 108 \\
$> 0.07$--$\leq 0.09$ & 41 & 10 & 50 & 47 & 10 & 59 \\
$> 0.09$--$\leq 0.12$ & 46 & 11 & 69 & 38 & 11 & 49 \\
$> 0.12$--$\leq 0.2$ & 43 & 10 & 60 & 53 & 5 & 59 \\
$> 0.2$ & 83 & 20 & 92 & 101 & 21 & 117 \\
\midrule
Total & 3337 & 3337 & 3337 & 3337 & 3337 & 3337 \\
\bottomrule
\end{tabular}
\end{table*}

\begin{table}[ht]
\caption{Three-model cross-tabulation after removing PLAW from the raw-rank comparison. The entries compare the model with the smallest reported reduced-statistic value among COMP, BAND, and SBPL on the FLNC and PFLX sides.}
\label{tab:noplaw-3x3}
\centering
\scriptsize
\begin{tabular}{lrrrr}
\toprule
FLNC/PFLX & pflx\_comp & pflx\_band & pflx\_sbpl & Row total \\
\midrule
flnc\_comp & 657 & 223 & 273 & 1153 \\
flnc\_band & 341 & 542 & 333 & 1216 \\
flnc\_sbpl & 334 & 272 & 403 & 1009 \\
Column total & 1332 & 1037 & 1009 & 3378 \\
\bottomrule
\end{tabular}
\end{table}

\clearpage

\section{Supplementary COMP and PLAW analyses}\label{app:compplawdetails}

\subsection{COMP reported fit-field comparison}\label{app:compfitfield}

The statistic-label and spectral-catalogue transition are identified in Sect.~\ref{subsec:catalogtransition}. Table~\ref{tab:compfit_fields} documents the corresponding numerical behaviour of the COMP \code{redchisq} and \code{redfitstat} fields. These results are presented for completeness, since the absolute fit-statistic analysis in Sect.~\ref{sec:compfitquality} is restricted to B.

\begin{table*}[htbp]
\centering
\scriptsize
\caption{Comparison of the reported COMP fit statistics. The interquartile range (IQR) column gives the 25th--75th percentile interval of \code{compredchisq}. Equality is the number of rows for which \code{compredchisq} and \code{compredfitstat} are exactly equal in the catalogue data.}
\label{tab:compfit_fields}
\begin{tabular}{llrrrrrr}
\toprule
Group & Spectrum & $N$ & Median \code{compredchisq} & IQR & Median redfitstat & Equal rows & Median DOF \\
\midrule
shortA & PFLX & 103  & 0.4660 & 0.4400--0.4950 & 0.4660 & 103/103 & 474 \\
shortA & FLNC & 103  & 0.4940 & 0.4435--0.5350 & 0.4940 & 103/103 & 446 \\
shortB & PFLX & 375  & 1.0190 & 0.9555--1.1045 & 0.9100 & 1/375 & 480 \\
shortB & FLNC & 375  & 1.0250 & 0.9730--1.0890 & 1.0460 & 1/375 & 480 \\
longA  & PFLX & 990  & 0.5430 & 0.5163--0.5690 & 0.5430 & 990/990 & 473 \\
longA  & FLNC & 990  & 0.5885 & 0.5430--0.7008 & 0.5885 & 990/990 & 473 \\
longB  & PFLX & 1848 & 1.0230 & 0.9708--1.0853 & 1.1000 & 3/1848 & 479 \\
longB  & FLNC & 1848 & 1.0730 & 1.0100--1.1580 & 1.3200 & 0/1848 & 479 \\
\bottomrule
\end{tabular}
\end{table*}

In every A row, the two reported COMP fields are numerically identical in both spectral blocks. In B, exact equality occurs in only one short PFLX row, one short FLNC row, three long PFLX rows, and no long FLNC row. Thus, the table gives the numerical support for treating the two catalogue periods separately in absolute fit-statistic analyses.

\subsection{COMP sensitivity to removing extreme fit-statistic values}\label{app:comprobustness}

We repeated the fixed-scale A--B physical comparisons after removing, within each group and relevant spectral block, either the upper 10\% of the reported COMP reduced-statistic distribution or the lowest and highest 5\% of the distribution. Table~\ref{tab:redchi_robustness} gives the results.

\begin{table*}[htbp]
\centering
\scriptsize
\caption{Sensitivity of the fixed-scale A--B COMP effects to within-group removal of extreme reported COMP reduced-statistic values.}
\label{tab:redchi_robustness}
\begin{tabular}{llrrrrr}
\toprule
Contrast & Removal & Variables & Median $|\Delta r|$ & Maximum $|\Delta r|$ & $N(\geq0.05)$ & $N(\geq0.10)$ \\
\midrule
A--B within short & Upper 10\% removed & 10 & 0.027 & 0.092 & 2 & 0 \\
A--B within short & Both 5\% tails removed  & 10 & 0.011 & 0.053 & 1 & 0 \\
A--B within long  & Upper 10\% removed & 10 & 0.012 & 0.042 & 0 & 0 \\
A--B within long  & Outer 5\% removed  & 10 & 0.008 & 0.027 & 0 & 0 \\
\bottomrule
\end{tabular}
\end{table*}

No A--B effect changes by 0.10 or more. In the omitted short--long FLNC comparisons, the largest absolute effect-size change is 0.037 in A and 0.066 in B.

Here we give the numerical PLAW tests needed to support Sect.~\ref{sec:plaw}.

\subsection{PLAW A-period amplitude-error field}\label{app:plawamperror}

In the A-period PLAW amplitude-error field we find a regular pattern: the error is approximately the fitted amplitude plus 0.1 in 61 of 103 shortA PFLX rows (59.2\%), 53 of 103 shortA FLNC rows (51.5\%), 121 of 990 longA PFLX rows (12.2\%), and 14 of 990 longA FLNC rows (1.4\%). None of these rows is catalogue PLAW-best, and the pattern is absent from B. Since the catalogue data do not identify its exact origin, we do not interpret the A-period amplitude error as an ordinary symmetric $1\sigma$ uncertainty.

\subsection{PLAW catalogue validation}\label{app:plawvalidation}

Table~\ref{tab:plaw_deltaS_check} compares the calculated $\Delta S_{\rm PC}>8.58$ rule with the A and B catalogue results.

\begin{table*}[htbp]
\centering
\scriptsize
\caption{PLAW$\rightarrow$COMP model comparison using the parameter-quality criterion. Agreement compares the decision $\Delta S_{\rm PC}>8.58$ with the catalogue non-PLAW result among rows for which both PLAW and COMP pass the criterion.}
\label{tab:plaw_deltaS_check}
\begin{tabular}{llrrrr}
\toprule
Group & Spectrum & Both models good $N$ & $\Delta S_{\rm PC}>8.58$ (\%) & Agreement (\%) & Median $\Delta S_{\rm PC}$ \\
\midrule
shortA & PFLX & 22 & 59.1 & 95.5 & 10.43 \\
shortA & FLNC & 36 & 86.1 & 100.0 & 81.33 \\
shortB & PFLX & 153 & 98.0 & 99.3 & 45.27 \\
shortB & FLNC & 239 & 97.5 & 99.6 & 58.63 \\
longA & PFLX & 728 & 68.0 & 99.0 & 14.43 \\
longA & FLNC & 919 & 86.2 & 99.6 & 47.12 \\
longB & PFLX & 1149 & 91.0 & 99.6 & 62.72 \\
longB & FLNC & 1589 & 93.7 & 99.2 & 92.22 \\
\bottomrule
\end{tabular}
\end{table*}

The calculated rule agrees with the catalogue result in 95.5--100\% of the tested rows. In B, 1494 of the 1497 PLAW-best cases are explained either by a COMP solution that fails the parameter-quality condition or by $\Delta S_{\rm PC}\leq8.58$; only three rows remain unexplained by this calculation. The raw-ranking results give the same conclusion as Sect.~\ref{subsec:plaw_strategy}: when all four models are well measured, PLAW is usually the poorest numerical fit, especially for FLNC.

\subsection{PLAW model dependence}\label{app:plawmodeldependence}

The main text shows that the PLAW index becomes more precise with brightness. Table~\ref{tab:plaw_comp_divergence} gives the PLAW--COMP index differences and the standardized rank-regression coefficient of $\Delta S_{\rm PC}$ for their absolute difference.

\begin{table*}[htbp]
\centering
\scriptsize
\caption{PLAW--COMP index differences in B. The final column is the standardized rank-regression coefficient of $\Delta S_{\rm PC}$ for the absolute index difference after brightness and DOF are controlled.}
\label{tab:plaw_comp_divergence}
\begin{tabular}{llrrrr}
\toprule
Group & Spectrum & $N$ & Median $(\lambda_{\rm PLAW}-\alpha_{\rm COMP})$ & Median absolute difference & $\Delta S_{\rm PC}$ coefficient [95\% CI] \\
\midrule
shortB & PFLX & 153 & -0.902 & 0.902 & 0.775 [0.485, 1.065] \\
shortB & FLNC & 239 & -0.798 & 0.798 & 0.451 [0.203, 0.699] \\
longB & PFLX & 1149 & -0.728 & 0.729 & 0.893 [0.801, 0.985] \\
longB & FLNC & 1589 & -0.580 & 0.581 & 1.024 [0.932, 1.117] \\
\bottomrule
\end{tabular}
\end{table*}

The disagreement with the COMP low-energy index increases with the evidence for curvature. Together with the brightness dependence of the PLAW index error reported in Sect.~\ref{subsec:plaw_brightness}, this result supports the interpretation of the PLAW index as an effective broad-band slope rather than as a model-independent physical low-energy index.

\subsection{PLAW robustness}\label{app:plawrobustness}

We repeated the principal A--B PLAW contrasts after removing extreme values of the reported PLAW reduced-statistic field within each group. Table~\ref{tab:plaw_robustness} gives the results.

\begin{table*}[htbp]
\centering
\scriptsize
\caption{Robustness of the principal A--B PLAW contrasts to within-group removal of extreme values of the reported PLAW reduced-statistic field.}
\label{tab:plaw_robustness}
\begin{tabular}{llrrrl}
\toprule
Contrast & Removal & Variables & Median $|\Delta r|$ & Maximum $|\Delta r|$ & Variable at maximum \\
\midrule
short A--B & upper 10\% removed & 6 & 0.028 & 0.041 & \code{pflxplawergflux} \\
short A--B & outer 5\% removed & 6 & 0.055 & 0.060 & \code{flncplawergflnc} \\
long A--B & upper 10\% removed & 6 & 0.003 & 0.008 & \code{pflxplawergflux} \\
long A--B & outer 5\% removed & 6 & 0.011 & 0.016 & \code{flncplawphtflnc} \\
\bottomrule
\end{tabular}
\end{table*}

No A--B effect changes by 0.10 or more. In the omitted short--long comparisons with extreme values removed, the largest changes are 0.062 in A and 0.056 in B. Much larger changes in brightness-related effects can appear when only PLAW-best events are selected because model selection itself depends strongly on brightness; we therefore do not use such subsets for the main physical comparisons. The A--B threshold tests are consistent with the main classifier result; adding the 21 residual FLNC rows to longB changes the tested PLAW effects by at most 0.0022.

\section{Supplementary BAND analyses}\label{app:banddetails}

Here we give the numerical BAND tests needed to support Sect.~\ref{sec:band} without repeating figures or additional classifier plots.

\subsection{Catalogue validation and sample reduction}

The full COMP--BAND catalogue comparison in B is given in Table~\ref{tab:band_ordering}.

\begin{table*}[htbp]
\centering
\scriptsize
\caption{Catalogue COMP--BAND model comparison in B using the conservative full parameter-quality criterion. Rows with an SBPL catalogue label are excluded from the two-model agreement denominator rather than assigned to COMP or BAND.}
\label{tab:band_ordering}
\begin{tabular}{llrrrr}
\toprule
Group & Spectrum & Both COMP and BAND good & $N(\dscb>11.83)$ & Fraction & Catalogue agreement \\
\midrule
shortB & PFLX & 9   & 4   & 44.4\% & 100.0\% \\
shortB & FLNC & 26  & 7   & 26.9\% & 100.0\% \\
longB  & PFLX & 275 & 105 & 38.2\% & 99.6\% \\
longB  & FLNC & 594 & 293 & 49.3\% & 98.8\% \\
\bottomrule
\end{tabular}
\end{table*}

The calculated $\dscb>11.83$ rule agrees with the catalogue COMP-versus-BAND result in 98.8--100\% of the tested B rows. The B FLNC medians also closely reproduce the published ten-year catalogue: $\alpha=-0.462$ and $-0.847$ and $E_{\rm peak}=420$ and 148 keV for short and long bursts, compared with the published $-0.46$, $-0.84$, 413, and 144 keV \citep{Poolakkil2021GBM10yr}. If a usable $\beta$ is required, the usable fraction is reduced to 1.9--37.5\% across the eight principal samples, whereas the $\alpha$--$E_{\rm peak}$ criterion retains 37.6--94.6\%.

\subsection{High-energy-index errors}

Table~\ref{tab:band_beta_errors} gives the detailed B comparison of the reported $\alpha$ and $\beta$ errors.

\begin{table*}[htbp]
\centering
\scriptsize
\caption{B comparison of the reported low- and high-energy-index errors.}
\label{tab:band_beta_errors}
\begin{tabular}{llrrrr}
\toprule
Group & Spectrum & Median $\sigma_{\alpha}$ & Median $\sigma_{\beta}$ & $\sigma_{\beta}>1$ & Exact $\beta=-2$ \\
\midrule
shortB & PFLX & 0.433 & 99.9  & 75.5\% & 0.27\% \\
shortB & FLNC & 0.262 & 12.6  & 69.1\% & 0.27\% \\
longB  & PFLX & 0.250 & 4.50  & 60.0\% & 0.65\% \\
longB  & FLNC & 0.132 & 0.490 & 42.6\% & 1.57\% \\
\bottomrule
\end{tabular}
\end{table*}

A also contains the strong exact-$\beta=-2$ concentration summarized in Sect.~\ref{subsec:band_strategy}, together with exact $\beta=-10$, reported errors of 8, and a few extreme $\alpha$ values. We do not interpret these patterns as continuous physical distributions.

\subsection{Precision and model dependence}

Table~\ref{tab:band_divergence} gives the COMP--BAND parameter-difference results.

\begin{table*}[htbp]
\centering
\scriptsize
\caption{Association between COMP--BAND fit improvement and model-dependent parameter differences in B. The coefficients are standardized rank-regression coefficients of $\dscb$ for the absolute COMP--BAND parameter differences after controlling for brightness and DOF; brackets give 95\% intervals.}
\label{tab:band_divergence}
\begin{tabular}{llrrr}
\toprule
Group & Spectrum & $N$ & $\dscb\rightarrow|\Delta\alpha|$ [95\% CI] & $\dscb\rightarrow|\Delta\log_{10}E_{\rm peak}|$ [95\% CI] \\
\midrule
shortB & PFLX & 134  & 0.411 [0.218, 0.603] & 0.544 [0.376, 0.712] \\
shortB & FLNC & 217  & 0.609 [0.474, 0.745] & 0.674 [0.543, 0.805] \\
longB  & PFLX & 1097 & 0.771 [0.716, 0.825] & 0.778 [0.724, 0.832] \\
longB  & FLNC & 1428 & 0.878 [0.833, 0.923] & 0.859 [0.812, 0.905] \\
\bottomrule
\end{tabular}
\end{table*}

In the highest brightness quintile, the $\alpha$--$E_{\rm peak}$ criterion is satisfied by 85.3\%, 97.3\%, 96.8\%, and 98.9\% of shortB PFLX, shortB FLNC, longB PFLX, and longB FLNC events, respectively. The corresponding full criterion including $\beta$ reaches only 17.3\%, 28.0\%, 60.3\%, and 67.8\%. Thus, brighter spectra often provide usable low-energy BAND parameters, whereas a reliable high-energy slope remains much less common.

\subsection{Robustness and classifier tests}

The omitted short--long rows reproduce the familiar short-hard and long-soft relation: the $\alpha$ effects are 0.524 in A and 0.561 in B, while the $\log E_{\rm peak}$ effects are 0.569 and 0.626. Removing the upper 10\% of the within-group BAND statistic changes no main effect by more than 0.0449, and adding the 21 residual FLNC rows to longB changes the tested effects by at most 0.0022. Full-$\beta$ and BAND-best restrictions can reduce the short cells to one or two events and therefore give unstable effects.

For duration classification, the BAND $\alpha+\log E_{\rm peak}$ model gives AUC 0.877 and balanced accuracy 0.802 in the 2735-event BAND shape sample. On the common BAND--COMP sample, the AUC values are 0.875 for BAND, 0.885 for COMP, and 0.884 for the combined shape. Training on A and testing on B gives AUC 0.870, while the reverse transfer gives 0.892. These results show that the BAND low-energy parameters can be examined in a separate variable set rather than combined with COMP as independent variables.

\section{Supplementary SBPL parameter-quality tests}\label{app:sbplquality}

The smoothly broken power-law fits provide a low-energy index $\lambda_1$, a break energy $E_{\rm b}$, and a high-energy index $\lambda_2$ \citep{smothly,2006ApJS..166..298K,gold12,gru14,yu_evolution,Poolakkil2021GBM10yr}. Here we test whether the two indices are measured with sufficiently similar accuracy for use in population studies. We use the following parameter-quality conditions:
\begin{align}
Q_1 &: 0<\sigma_{\lambda_1}\leq0.4,\quad \lambda_1>-2,\quad |\lambda_1|<5,\\
Q_2 &: 0<\sigma_{\lambda_2}\leq0.4,\quad -10<\lambda_2<-2,\\
Q_{\rm b} &: E_{\rm b}>0,\quad 0<\left|\frac{\sigma_{E_{\rm b}}}{E_{\rm b}}\right|\leq0.4.
\end{align}
These broad value limits remove obvious boundary or extreme solutions and do not define a physical population. A $\lambda_2$ entry is separately flagged as boundary/extreme or zero-error when $\sigma_{\lambda_2}=0$, $\lambda_2\leq-10$, or $\lambda_2\geq-2$. PFLX and FLNC are kept separate, and the PFLX rows are interpreted only within their fixed 64 or nominal 1024 ms groups.

Across all 3337 rows, we find that $Q_1$ is satisfied by 2515 PFLX fits (75.37\%) and 2878 FLNC fits (86.25\%). The corresponding $Q_2$ counts are only 688 (20.62\%) and 1260 (37.76\%), while the full three-parameter criterion is satisfied by 580 PFLX fits (17.38\%) and 1144 FLNC fits (34.28\%). Table~\ref{tab:sbpl_quality} shows larger differences among the four principal group--spectrum samples: the fraction of usable $\lambda_2$ fits ranges from 3.7\% in shortB PFLX to 53.1\% in longA FLNC, and the boundary/extreme-or-zero-error fraction ranges from 26.4\% to 47.6\%.

The low-energy index nevertheless contains consistent spectral information. Table~\ref{tab:sbpl_agreement} compares $\lambda_1$ with the COMP low-energy index and BAND $\alpha$ after applying the corresponding parameter-specific error conditions. The groupwise Spearman correlations are 0.765--0.854 against COMP and 0.665--0.921 against BAND. Hence, a well-measured $\lambda_1$ follows the same broad low-energy hardness direction as the other curved models.

The usable $\lambda_1$ subsets still show a brightness selection: their median brightness is 1.66--6.80 times that of the non-usable subsets across the eight principal samples. Since usable measurements occur mainly among brighter bursts, $\lambda_1$ cannot be used as a common parameter for the full catalogue despite its consistent low-energy hardness information.

The brightness selection is stronger for $\lambda_2$: usable subsets are 2.83--5.24 times brighter than non-usable subsets. A well-measured $\lambda_2$ can agree closely with BAND $\beta$ (longB: $\rho=0.916$, $N=232$, in PFLX; $\rho=0.882$, $N=462$, in FLNC), but the overlap is absent in shortA and small in the other short and A PFLX samples. The differences in the fractions of usable measurements, boundary or extreme values, zero reported errors, long error tails, and brightness selection mean that $\lambda_2$ cannot be treated as a reliable parameter across the full catalogue.

The results therefore support a limited interpretation: $\lambda_1$ retains useful low-energy hardness information when well measured, whereas $\lambda_2$ requires verification for an individual burst and is not suitable for general population studies of the full catalogue.

\onecolumn
\begin{table}[htbp]
\centering
\scriptsize
\caption{SBPL parameter quality in the four principal catalogue-period and PFLX-scale groups. ``Usable'' follows $Q_1$, $Q_2$, or $Q_{\rm b}$ above. ``Flagged $\lambda_2$'' is the union of zero reported error, $\lambda_2\leq-10$, and $\lambda_2\geq-2$. ``Full shape'' requires $Q_1\cap Q_2\cap Q_{\rm b}$. Counts are followed by percentages of the corresponding row. The 21 non-standard-interval \code{unic} events are omitted because they do not form a standard PFLX group.}
\label{tab:sbpl_quality}
\resizebox{\textwidth}{!}{%
\begin{tabular}{llrrrrrrrr}
\toprule
Group & Spectrum & $N$ & Median $\sigma_{\lambda_1}$ & Usable $\lambda_1$ & Median $\sigma_{\lambda_2}$ & Usable $\lambda_2$ & Flagged $\lambda_2$ & Usable $E_{\rm b}$ & Full shape \\
\midrule
\code{shortA} & PFLX & 103 & 0.259 & 72 (69.9\%) & 0.381 & 13 (12.6\%) & 48 (46.6\%) & 86 (83.5\%) & 12 (11.7\%) \\
\code{shortA} & FLNC & 103 & 0.165 & 86 (83.5\%) & 0.241 & 20 (19.4\%) & 49 (47.6\%) & 91 (88.3\%) & 18 (17.5\%) \\
\code{shortB} & PFLX & 375 & 0.325 & 217 (57.9\%) & 3.600 & 14 (3.7\%) & 169 (45.1\%) & 68 (18.1\%) & 8 (2.1\%) \\
\code{shortB} & FLNC & 375 & 0.212 & 281 (74.9\%) & 1.820 & 30 (8.0\%) & 119 (31.7\%) & 121 (32.3\%) & 24 (6.4\%) \\
\code{longA} & PFLX & 990 & 0.173 & 836 (84.4\%) & 0.225 & 296 (29.9\%) & 384 (38.8\%) & 782 (79.0\%) & 259 (26.2\%) \\
\code{longA} & FLNC & 990 & 0.079 & 924 (93.3\%) & 0.137 & 526 (53.1\%) & 261 (26.4\%) & 912 (92.1\%) & 511 (51.6\%) \\
\code{longB} & PFLX & 1848 & 0.200 & 1372 (74.2\%) & 1.300 & 360 (19.5\%) & 604 (32.7\%) & 729 (39.4\%) & 297 (16.1\%) \\
\code{longB} & FLNC & 1848 & 0.102 & 1569 (84.9\%) & 0.356 & 675 (36.5\%) & 521 (28.2\%) & 1132 (61.3\%) & 584 (31.6\%) \\
\bottomrule
\end{tabular}%
}
\end{table}

\begin{table}[htbp]
\centering
\scriptsize
\caption{Agreement and brightness selection of the SBPL indices. $N_{1C}$ and $N_{1B}$ are the overlaps used for the Spearman correlations of SBPL $\lambda_1$ with the COMP index and BAND $\alpha$, respectively. $B_1$ and $B_2$ are the ratios of median brightness in the usable and non-usable $\lambda_1$ or $\lambda_2$ subsets. Brightness is \code{flux64} for short PFLX, \code{flux1024} for long PFLX, and catalogue fluence for FLNC. See text for the interpretation of the brightness-selection ratios.}
\label{tab:sbpl_agreement}
\begin{tabular}{llrrrrrr}
\toprule
Group & Spectrum & $N_{1C}$ & $\rho(\lambda_1,\alpha_{\rm COMP})$ & $N_{1B}$ & $\rho(\lambda_1,\alpha_{\rm BAND})$ & $B_1$ & $B_2$ \\
\midrule
\code{shortA} & PFLX & 69 & 0.804 & 64 & 0.801 & 2.79 & 4.61 \\
\code{shortA} & FLNC & 85 & 0.765 & 78 & 0.665 & 6.80 & 4.62 \\
\code{shortB} & PFLX & 171 & 0.845 & 129 & 0.921 & 1.66 & 2.90 \\
\code{shortB} & FLNC & 259 & 0.854 & 210 & 0.902 & 2.08 & 5.24 \\
\code{longA} & PFLX & 833 & 0.777 & 754 & 0.673 & 1.76 & 2.83 \\
\code{longA} & FLNC & 920 & 0.802 & 885 & 0.776 & 4.16 & 3.51 \\
\code{longB} & PFLX & 1222 & 0.847 & 1069 & 0.908 & 1.89 & 4.47 \\
\code{longB} & FLNC & 1550 & 0.776 & 1377 & 0.881 & 2.83 & 4.12 \\
\bottomrule
\end{tabular}
\end{table}

\end{appendix}
\end{document}